\documentclass[
  reprint,
  amsmath,amssymb,
  aps,
  pra,
  floatfix,
  nofootinbib
]{revtex4-2}

\usepackage{needspace}
\usepackage{ifthen}
\usepackage{diagbox}
\usepackage{setspace}
\usepackage{caption}
\usepackage[labelsep=period]{caption}
\usepackage{quantikz}
\usepackage{float}
\usepackage{amsthm}
\usepackage{graphicx}
\usepackage{booktabs}
\usepackage{makecell}
\usepackage{dcolumn}
\usepackage{multirow}
\usepackage{bm}
\usepackage{comment}
\usepackage{algorithm}
\usepackage{algpseudocode}
\usepackage{physics}
\usepackage{enumitem}
\usepackage{threeparttable}
\usepackage{orcidlink}
\usepackage{xcolor}
\usepackage{hyperref}
\hypersetup{colorlinks = true, linkcolor=blue, citecolor = blue, urlcolor=blue}

\newtheoremstyle{customdef}
  {3pt}{3pt} 
  {}        
  {}         
  {\itshape} 
  {}      
  { }
  {\indent\thmname{#1}~\thmnumber{#2}.{\ }{\normalfont(\upshape #3)}}
\theoremstyle{customdef}
\newtheorem{definition}{Definition}
\newtheorem{theorem}{Theorem}
\newtheorem{lemma}{Lemma}
\newtheorem{corollary}{Corollary}
\newtheorem{Proof of Theorem}{Proof of Theorem}

\newcounter{algnum}
\newenvironment{PRAalgorithm}[2][]{%
  \refstepcounter{algnum}%
  \noindent\textbf{ALGORITHM \thealgnum. #1}\label{#2}\par
  \vspace{0.3ex}
  \hrule height 1pt
  \vspace{1pt}
  \hrule height 1pt
  \vspace{2pt}
  \begin{algorithmic}[1]
}{%
  \end{algorithmic}
  \vspace{2pt}
  \hrule height 1pt
  \vspace{1pt}
  \hrule height 1pt
  \par\vspace{3pt}
}

\makeatletter
\newif\ifreprint
\@ifclasswith{revtex4-2}{reprint}{\reprinttrue}{\reprintfalse}
\makeatother

\makeatletter
\newcounter{stepsection}

\newcommand{\stepsection}[1]{%
    \refstepcounter{stepsection}%
    \begin{center}
      \textbf{\textit{\thestepsection.~#1}}%
    \end{center}
}
\makeatother

\makeatother

\begin{document}
\preprint{APS/123-QED}

\title{Optimizing sparse quantum state preparation with measurement and feedforward}

\author{Yao-Cheng Lu\orcidlink{0009-0007-6024-118X}}
\altaffiliation[Contact author: ]{\href{mailto:yao.cheng.lu@gapp.nthu.edu.tw}{yao.cheng.lu@gapp.nthu.edu.tw}}
\author{Han-Hsuan Lin\orcidlink{0000-0002-5126-0174}}
\altaffiliation[Contact author: ]{\href{mailto:linhh@cs.nthu.edu.tw}{linhh@cs.nthu.edu.tw}}
\affiliation{
 Department of Computer Science, National Tsing Hua University, Hsinchu 300044, Taiwan
}

\begin{abstract}
    
Quantum state preparation (QSP) is a key component in many quantum algorithms. In particular, the problem of sparse QSP (SQSP) — the task of preparing the states with only a small number of non-zero amplitudes — has garnered significant attention in recent years. In this work, we focus on reducing the circuit depth of SQSP with limited number of ancilla qubits. We present two SQSP algorithms: one with depth $O(n\log d)$, and another that reduces depth to $O(n)$. The latter leverages mid-circuit measurement and feedforward, where intermediate measurement outcomes are used to control subsequent quantum operations. Both constructions have size $O(dn)$ and use $O(d)$ ancilla qubits. Compared to the state-of-the-art SQSP algorithm in  \href{https://ieeexplore.ieee.org/abstract/document/10044235}{Sun et al., IEEE, 2023}, which allows an arbitrary number of ancilla qubits $m>0$, both of our algorithms achieve lower circuit depth when $m=d$.
\end{abstract}

\maketitle

\section{\label{sec:1}Introduction}

\subsection{Motivation\label{sec:1.1}}

Since the advent of quantum computing, numerous quantum algorithms have been developed, demonstrating significant speedups over classical algorithms.~\cite{Shor_1996,Grover_1996} A fundamental step in many  algorithms — such as variational quantum algorithms~\cite{VA_2020}, hybrid quantum-classical Monte-Carlo methods~\cite{MC_2018,MC_2022}, and quantum linear solvers~\cite{QLSS_2024,QLSS_2023} — is the initialization of classical data on a quantum device. Furthermore, some algorithms~\cite{CNN} may require reinitializing the quantum state multiple times during execution. Quantum state preparation is a particularly important step in these algorithms and can often dominate the overall run time. Therefore, efficiently preparing quantum states is a critical challenge in quantum computing.

The asymptotically optimal bounds for the depth complexity of the general quantum state preparation (GQSP) without ancilla qubits is $\Theta\left(\frac{2^n}{n}\right)$~\cite{ODQSP_2021}. When ancilla qubits are available, the trade-offs depend on the number of ancilla qubits used. With $m$ ancilla qubits, if $m = O(\frac{2^n}{n\log n})$, the circuit depth can be reduced to $\Theta(\frac{2^n}{n + m})$. If $m=\Omega(2^n)$, the circuit depth can be further reduced to $\Theta(n)$. Both bounds asymptotically optimal~\cite{ODQSP_2021}. When the number of ancilla qubits is within the range $\omega(\frac{2^n}{n\log n})$ to $o(2^n)$, the circuit depth ranges between $\Omega(n)$ and $O(n\log n)$~\cite{ODQSP_2021}. All GQSP algorithms require a circuit size of $\Theta(2^n)$, which demands substantial computational resources~\cite{GQSP_LB}.

Due to the high cost of preparing a general quantum state, several works have focused on designing QSP algorithms for states with specific structures that arise from particular applications, such as uniform quantum states~\cite{UQSP_2023}, low-rank quantum states~\cite{LRQSP_2021}, and cyclic quantum states~\cite{CQSP_2022}. Leveraging these
structures allows for a more efficient generation of quantum states. The sparse quantum states, quantum states with few non-zero amplitudes, are another category of quantum states with specific structure and practical applications. Bell states~\cite{BELLS_1998, GHZS_2018}, thermofield double states~\cite{TFDS_2018}, W states~\cite{WQSP_2000}, Dicke states~\cite{DQSP_2019}, and GHZ states~\cite{GHZS_2018} all fall into this category. Sparse quantum states have applications in several algorithms, such as quantum Byzantine agreement algorithm~\cite{RBAA_1985}. The problem of sparse quantum state preparation (SQSP) has been studied extensively recently~\cite{SQSP_2021,DDQSP_2022,EQSP_2018,SQSP_2023,DQSP_2021,ODQSP_2022,SQSP_2024,CQSP_2020,QW_2024,ODQSP_2021,CCQSP_2024}. In this work, we focus on solving the SQSP problem, aiming to minimize the resources required.

\begin{definition}[SQSP Problem]
Given two positive integers $n$ and $0<d\leq2^n$, and a set $S$ consisting of $d$ tuples:

\begin{equation}
    S=\{(\alpha_i,q_i)\; |\; \alpha_i\in \mathbb{C},q_i\in\{0,1\}^n,0\leq i< d\},
\end{equation}
where $\sum_{i=0}^{d-1}\lvert\alpha_i\rvert^2=1$. We define an $n$-qubit $d$-sparse state $\ket{\phi(n,d,S)}$ as below:
\begin{equation}
\ket{\phi(n,d,S)}=\sum_{i=0}^{d-1}\alpha_i\ket{q_i}.
\end{equation}
Sometime we abbreviate $\ket{\phi(n,d,S)}$ as $\ket{\phi}$. The goal of the SQSP problem is to generate an arbitrary sparse quantum state given $n,\,d,\,S$.
\end{definition}

In the Noisy Intermediate-Scale Quantum (NISQ) era, both the circuit depth and the number of qubits play the critical roles in determining the feasibility of a quantum algorithm. Limited coherence time constrains the achievable circuit depth, while fewer qubits simplify hardware control and reduce error rates. In many SQSP algorithms, circuit depth compression relies on the use of parallel operations, which typically require additional ancilla qubits. However, this space-for-time trade-off can become impractical for target states with large $n$, where the number of ancilla qubits may scale linearly or more with $n$, making it difficult to balance circuit depth and qubit overhead.

To address this challenge, our first construction keeps the number of ancilla qubits at $O(d)$, which can be significantly lower than many prior algorithms, while maintaining circuit depth $O(n\log d)$. This offers a practical balance between depth and qubit usage, making our method particularly suitable for the case with limited qubit resources, especially when $n\gg d$, where either circuit depth or ancilla used can be significantly reduced compared to known algorithms (See Table~\ref{tab:Table1}).

Furthermore, building upon the first algorithm, we incorporate measurement and feedforward to further reduce the circuit depth. By leveraging intermediate measurement outcomes to control subsequent quantum operations, our $\mathsf{MaF}$-based construction achieves circuit depth $O(n)$ that is independent of the sparsity $d$.

\subsection{Measurement and feedforward\label{sec:1.2}}

Measurement and feedforward is a quantum circuit design architecture that enables depth reduction by integrating mid-circuit measurements and classical feedforward to guide quantum operations. We denote $\mathsf{MaF}(\mathcal{Q}, \mathcal{C}, l)$ as the class of circuit with measurement and feedforward consisting of $l$ alternating quantum and classical layers. Each quantum layer is a quantum circuit in the class $\mathcal{Q}$, and each classical layer is a classical circuit in the class $\mathcal{C}$. In this structure, quantum layers apply unitary operations and mid-circuit measurements on selected qubits. When mid-circuit measurements are performed, the measured qubits collapse to a definite classical state, while unmeasured qubits remain unaffected. The measurement results are first processed classically, and then used to apply conditional gates that control subsequent quantum operations. This dynamic adaptation enables circuits to adjust during execution, making quantum computations more flexible. In this work, we use $\mathsf{MaF}$ to abbreviate $\mathsf{MaF}(\mathrm{QPoly(n)}, \mathrm{NC}^1, n)$.\footnote{QPoly(n) denotes the class of quantum circuits with polynomial size. We abuse the notation by using $\mathrm{NC}^1$  to denote the class of classical circuit with logarithmic depth and polynomial size, rather than the corresponding class of decision problems.} We refer to a circuit within this framework as a $\mathsf{MaF}$ circuit. A formal definition of $\mathsf{MaF}$ will be provided in Chapter~\ref{sec:2.2}.

$\mathsf{MaF}$ has a wide range of applications in quantum computing. For example, in quantum error correction~\cite{QEC}, measurement outcomes determine the corrective steps needed to maintain fault tolerance. Similarly, in quantum teleportation~\cite{Tel}, measurement outcomes are classically transmitted to a distant party, enabling reconstruction of the original quantum state. Furthermore, dynamic circuits with $\mathsf{MaF}$ have shown significant potential in the preparation of long-range entangled states~\cite{MaFLR}.

In this work, the $\mathsf{MaF}$-based design plays a crucial role in achieving efficient parallel execution. The algorithm involves extensive use of fan-out gates to enable parallel operations and thereby speed up the circuit. By implementing these fan-out gates with a $\mathsf{MaF}$ circuit, their depth can be minimized from logarithmic to constant with respect to the number of qubits of the target state. As a result, the overall circuit depth of the algorithm is reduced and made independent of the sparsity $d$.

\subsection{\label{sec:1.3}Contributions and comparison}
\begingroup
\setlength{\tabcolsep}{0.3cm}

In this work, we study the SQSP algorithms and focus on optimizing the circuit depth and the number of ancilla qubits.
We study two cases: without and with using a $\mathsf{MaF}$ circuit.

First, we give a construction without a $\mathsf{MaF}$ circuit:
\begin{theorem}[SQSP algorithm]
\label{th:1}
With the gate set drawn from $\text{\{U(2),CNOT}\}$, arbitrary $n$-qubit $d$-sparse quantum states can be prepared with a circuit of size $O(dn)$, depth $O(n\log d)$, and $O(d)$ ancilla qubits.
\end{theorem}

Compared to existing SQSP algorithms that use a constant number of  ancilla qubits~\cite{SQSP_2021,DDQSP_2022,EQSP_2018,SQSP_2023,DQSP_2021,QW_2024, QW_2024, CCQSP_2024, CQSP_2020, SQSP_2024}, our algorithm achieves a lower circuit depth of $O(n \log d)$. While some SQSP algorithms attain lower circuit depth, such as~\cite{ODQSP_2022} and ~\cite{CCQSP_2024}, they require a much larger number of ancilla qubits\textemdash $O(dn \log d)$ and $O(\frac{dn}{\log d})$, respectively\textemdash which can in turn increase the total qubit requirements of the overall algorithm. In contrast, our algorithm achieves a reasonable circuit depth while using only $O(d)$ ancilla qubits. Finally, although the algorithm in \cite{ODQSP_2021} supporting a trade-off with $m$ ancilla qubits for any $m\geq0$, our algorithm still achieves a lower circuit depth under the condition $m = O(d)$.

\begin{table*}[ht]
    \centering
    \renewcommand\arraystretch{1.8}
    \tabcolsep= 0.4 cm
    \resizebox{0.95\textwidth}{!}{
    \begin{threeparttable}
    \caption{Comparison of our result to previous SQSP methods without a $\mathsf{MaF}$ circuit.}
    \vspace{0.3cm}
    \label{tab:Table1}
    \begin{tabular}{ccccc}
        \toprule  \hline
        Algorithm & Circuit Size & Depth & \#Ancilla \\ \hline
        \makecell[c]{\cite{SQSP_2021, QW_2024}} & $O(dn)$ & $O(dn)$ & $0$ \\
        \makecell[c]{\cite{CCQSP_2024}} & $O\left(\frac{dn}{\log n}+n\right)$ & $/$ & $0$ \\
        \makecell[c]{\cite{DDQSP_2022,EQSP_2018,SQSP_2023,DQSP_2021}} & $O(dn)$ & $O(dn)$ & $1$  \\
        \makecell[c]{\cite{SQSP_2024}} & $O\left(\frac{dn}{\log n}+n\right)$ & $/$ & $2$  \\
        \makecell[c]{\cite{CQSP_2020}} & $O(dn)$ & $/$ & $2$  \\
        \makecell[c]{\cite{CCQSP_2024}} & $O\left(\frac{dn}{\log d}\right)$ & $O(\log dn)$ & $O\left(\frac{dn}{\log d}\right)$ \\
        \makecell[c]{\cite{ODQSP_2022}} & $O(dn\log d)$ & $\Theta(\log dn)$ & $O(dn\log d)$  \\
        \makecell[c]{\cite{ODQSP_2021}} & $/$ & $O(n\log dn +\frac{dn^2\log d}{n+m})$ & $O(m)$\\
        \makecell[c]{\cite{CCQSP_2024}} & $\Omega\left(\frac{dn}{\log (m+n)+\log d}+n\right)$ & $/$ & $O(m)\tnote{a}$ \\
        \makecell[c]{\cite{CCQSP_2024}} & $\Theta\left(\frac{dn}{\log dn}+n\right)$ & $/$ & unlimited number \\
        Theorem~\ref{th:1} & $O(dn)$ & $O(n\log d)$ & $O(d)$ \\  \hline
        \bottomrule
    \end{tabular}
    \begin{tablenotes}
    \item[a] under the condition $m=O(dn)$
    \end{tablenotes}
    \end{threeparttable}
    }
\end{table*}

While Theorem \ref{th:1} presents an efficient construction without a $\mathsf{MaF}$ circuit, integrating $\mathsf{MaF}$-based design into the construction allows us to achieve an even lower circuit depth. This leads to the following improved result:
\begin{theorem}[SQSP algorithm with a $\mathsf{MaF}$ circuit]

\label{th:2}
With the gate set drawn from $\text{\{U(2),CNOT\}}$, arbitrary $n$-qubit $d$-sparse quantum states can be prepared using a circuit of size $O(dn)$, depth $O(n)$, and $O(d)$ acilla qubits within the class $\mathsf{MaF}(\mathrm{QPoly(n)}, \mathrm{NC}^1, n)$.
\end{theorem}

\begin{table}[ht]
    \renewcommand\arraystretch{1.8}
    \captionsetup{justification=raggedright, singlelinecheck=false}
    \caption{Comparison of our result to previous SQSP method with a $\mathsf{MaF}$ circuit.}
    \label{tab:Table2}
    \centering
    \begin{tabular}{ccccc}
         \toprule  \hline
        Algorithm & Circuit Size & Depth & \#Ancilla \\ \hline
        \makecell[c]{\cite{CDSQSP_2025}} & $O(d^2\log n)$ & $\Theta(1)$ & $O(d^2\log n)$ \\
        Theorem~\ref{th:2} & $O(dn)$ & $O(n)$ & $O(d)$ \\  \hline
        \bottomrule
    \end{tabular}
\end{table}

Compared to the constant-depth SQSP algorithm~\cite{CDSQSP_2025}, our approach significantly reduces both the circuit size and the number of ancilla qubits. Moreover, since many algorithms using GQSP as a subroutine have $O(n)$ circuit depths~\cite{LD_QFT, LD_QTDA}, an SQSP algorithm with linear depth remains competitive in these cases.

The results of the two algorithms, along with comparisons to existing algorithms, are presented in Table~\ref{tab:Table1} and Table~\ref{tab:Table2}, respectively.
\endgroup

\subsection{\label{sec:1.4}Overview of algorithm}

A SQSP algorithm can be implemented through various approaches. Some methods set computational basis states into correct amplitudes one by one by applying many multi-controlled single-target gates~\cite{SQSP_2024}, while others employ decision trees to perform the correct rotations on each of the $n$ qubits sequentially~\cite{DDQSP_2022}. In our work, we adopt a permutation-based strategy, as also employed in \cite{SQSP_2023}. The core idea is to initially produce the correct amplitudes with $\lceil\log d\rceil$ qubits, and then move each amplitude to its desired position. Below is a high-level overview of the algorithm. Let $\ket{\phi}=\sum_{i=0}^{d-1}\alpha_i\ket{q_i}$ be the target quantum state. The first step uses the GQSP algorithm to prepare the state ${\ket{\phi'}}$:
\begin{equation}
    \ket{\phi'} = \sum_{i=0}^{d-1}\alpha_i\ket{i},
\end{equation}
whose amplitudes correspond to the non-zero entries of $\ket{\phi}$ using only $\lceil\log d\rceil$ qubits. In the next step, we first expand $\ket{\phi'}$ to $n$ qubits, and then apply a permutation unitary that maps $\ket{\phi'}$ to $\ket{\phi}$. We can split this step into three individual steps: one-hot encoding, permutation and garbage elimination.

In the one-hot encoding step, we project $d$ distinct entries onto $d$ ancilla qubits:

\begin{equation}
    \sum_{i=0}^{d-1}\alpha_i\ket{i}\ket{0}^{\otimes n-\lceil \log d\rceil}\ket{0}^{\otimes d} \longrightarrow \sum_{i=0}^{d-1}\alpha_i\ket{i}\ket{0}^{\otimes n-\lceil \log d\rceil}\ket{e_i},
\end{equation}
where $\ket{e_i}$ denotes the quantum state in which the $i$th qubit is $\ket{1}$, while all other qubits are $\ket{0}$, known as the one-hot encoding of $i$. This allows next step to be executed in parallel, thereby minimizing circuit depth.

Next, we are entering the most pivotal phase known as permutation step. The purpose of this step is to position each amplitude in its appropriate position, in other words, to transform $\ket{i}$ to $\ket{q_i}$:

\begin{equation}
    \sum_{i=0}^{d-1}\alpha_i\ket{i}\ket{0}^{\otimes n-\lceil \log d\rceil}\ket{e_i} \longrightarrow \sum_{i=0}^{d-1}\alpha_i\ket{q_i}\ket{e_i}.
\end{equation}

The $n$ qubits of $\ket{\phi'}\ket{0}^{\otimes n-\lceil \log d\rceil}$ are processed in sequence. By controlling specific qubits of $\ket{e_i}$, we can determine whether each qubit will be flipped. To process all $d$ basis states in parallel, we use a quantum OR-controlled single target X gate controlled by the one-hot state. The target qubit is flipped if any of the control qubits is $\ket{1}$. After completing this step, the desired state $\ket{\phi}$ is now prepared in the first $n$ qubits. Finally, the ancilla qubits need to be reset to 0. The circuit for executing this algorithm requires size $O(dn)$, depth $O(n\log d)$, and $O(d)$ ancilla qubits.

To further reduce the circuit depth, we adopt the Measurement and Feedforward ($\mathsf{MaF}$) design~\cite{shallow_2024}. It is a circuit design architecture that enables depth reduction by integrating mid-circuit measurements and classical feedforward to guide quantum operations. We adopt the constant-depth construction of arbitrary qubit fan-out gates from~\cite{LAQCC_2024}. The fan-out gates used in the one-hot encoding step and garbage elimination step can be optimized using the constant-depth construction, thereby reducing the circuit depth. In the permutation step, due to the property established in Lemma \ref{Lm:2}, the quantum OR-controlled X gate can be replaced by a parity-controlled X gate, which can be implemented using a fan-out gate and Hadamard gates. As a result, all three steps can leverage the MaF to reduce depth, leading to the result presented in Theorem \ref{th:2}.
\begin{lemma}[Property of parity-controlled X gate]
\label{Lm:2}
Under the condition that the input is a one-hot state, the parity-controlled X gate has the same effect as the quantum OR-controlled X gate.
\end{lemma}

\begin{table*}[t]
    \renewcommand\arraystretch{1.65}
    \tabcolsep=0.5cm
    \vspace{0.1cm}
    \caption{Complexity summary for each step of our work.}
    \label{tab:Table3}
    \begin{center}
    \resizebox{0.95\textwidth}{!}{
    \begin{tabular}{ccccc}
        \hline \hline
        \multirow{2}{*}{\diagbox{\hspace{10ex}Step}{\hspace{-10ex}Complexity}} & \multirow{2}{*}{\raisebox{-4.5ex}{Size}} & \multicolumn{2}{c}{Depth} & \multirow{2}{*}{\raisebox{-4.5ex}{\#Ancilla}} \\
        \cline{3-4}
        & & \centering Without $\mathsf{MaF}$ & \centering With $\mathsf{MaF}$ & \\
        \hline
        GQSP & $O(d)$ & $O(\log d)$ & $O(\log d)$ & $O(d)$ \\
        One-hot encoding & $O(d)$ & $O(\log^2 d)$ & $O(\log d)$ & $O(d)$ \\
        Permutation & $O(dn)$ & $O(n\log d)$ & $O(n)$ & $O(d)$ \\
        Garbage elimination & $O(d)$ & $O(n\log d)$ & $O(n)$ & $O(d)$ \\
        Total & $O(dn)$ & $O(n\log d)$ & $O(n)$ & $O(d)$ \\
        \hline \hline
    \end{tabular}
    }
    \end{center}
\end{table*}

Table~\ref{tab:Table3} presents the circuit size, depth, and number of ancilla qubits required to execute the four steps of the algorithms, comparing both the cases with and without $\mathsf{MaF}$. Due to the fact that $d \leq 2^n$, we can regard $O(\log d)$ as $O(n)$. In summary, both algorithms can be executed with the same size and number of ancilla qubits, $O(dn)$ and $O(d)$, respectively. The algorithm without $\mathsf{MaF}$ has a circuit depth of $O(n\log d)$. However, the depth can be reduced to a linear complexity in $n$ by employing the $\mathsf{MaF}$ framework.

In Chapter \ref{chap:pre}, we briefly introduce the required quantum gates, $\mathsf{MaF}$, and then review the GQSP algorithm. The main algorithm is explained in detail in Chapter \ref{chap:SQSP}. Finally, we discuss future work and summarize our research in Chapter \ref{chap:ccl}.

\section{\label{chap:pre}Preliminaries}
\subsection{Quantum gates}

In this work, we will use the folowing gates: X gate, Hadamard gate, $\text{R}_y$ gate, Controlled-NOT gate, Controlled-$\text{R}_y$ gate, Fredkin gate, Toffoli gate, multi-controlled single-target gate, and quantum fan-out gate.

\begin{table}[H]
    \renewcommand\arraystretch{1.7}
    \tabcolsep= 0.9 cm
    \caption{Elementary Quantum Gates.}
    \vspace{0.3ex}
    \label{tab:Table4}
    \centering
    \resizebox{\columnwidth}{!}{
    \begin{tabular}{lccc}
        \toprule  \hline
        \makecell[c]{\large \textbf{Operator}} & \large\textbf{Gate} & \large\textbf{Unitary} \\
        \midrule
         \makecell[c]{\large  \textbf{Pauli-X (X)}} & \begin{quantikz} &\gate{\text{X}} & \end{quantikz} & $ \left( \begin{array}{cc} 0 & 1 \\ 1 & 0 \\ \end{array} \right)$ \\  
         \makecell[c]{\large \textbf{Hadamard (H)}} & \begin{quantikz} &\gate{\text{H}} & \end{quantikz} & $ \frac{1}{\sqrt{2}}\left( \begin{array}{cc} 1 & 1 \\ 1 & -1 \\ \end{array} \right)$ \\
         
         \makecell[c]{\large \textbf{Y-axis Rotation ($R_y(\theta)$)}} & \begin{quantikz} &\gate{\text{$\text{R}_y(\theta)$}} & \end{quantikz} & $\left( \begin{array}{cc} \cos{\frac{\theta}{2}} & -\sin{\frac{\theta}{2}} \\ \sin{\frac{\theta}{2}} & \cos{\frac{\theta}{2}} \\ \end{array} \right)$ \\

         \makecell[c]{\large \textbf{Controlled Not (CNOT)}} &\begin{quantikz} & \ctrl{1} & \\ & \targ{} & \end{quantikz} & $\left( \begin{array}{cccc} 1 & 0 & 0 & 0\\ 0 & 1 & 0 & 0\\ 0 & 0 & 0 & 1 \\ 0 & 0 & 1 & 0\\ \end{array}\right)$ \\
         
         \makecell[c]{\large \textbf{$\pi/8$ (T)}} &\begin{quantikz} & \gate{\text{T}} & \end{quantikz} & $\left( \begin{array}{cc} 1 & 0 \\ 0 & e^{\frac{i\pi}{4}} \end{array}\right)$ \\
         \hline
        \bottomrule
    \end{tabular}
    }
\end{table}

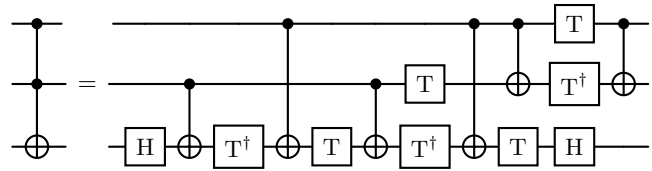
\begin{figure}[H]
    \centering
    \begin{quantikz}[column sep=1.2ex, row sep=2ex]
    & \ctrl{2} &  \midstick[3,brackets=none]{=} &&&&\ctrl{2}&&&&\ctrl{2}&\ctrl{1}&\gate{\text{T}}&\ctrl{1}&\\
    & \ctrl{0} & &&\ctrl{1}&&&&\ctrl{1}&\gate{\text{T}}&&\targ{}&\gate{\text{T}^\dag}&\targ{}&\\
    & \targ{} & & \gate{\text{H}} & \targ{} & \gate{\text{T}^\dag}&\targ{}&\gate{\text{T}}&\targ{}&\gate{\text{T}^\dag}&\targ{}&\gate{\text{T}}&\gate{\text{H}}&& 
    \end{quantikz}
    \vspace{0.2cm}
    \caption{Toffoli Gate}
    \label{fig:Tof}
\end{figure}

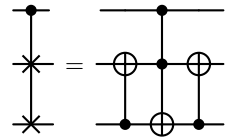
\begin{figure}[H]
    \centering
    \begin{quantikz}[column sep=1.2ex, row sep=3.5ex]
    &\ctrl{2} & \midstick[3,brackets=none]{=} &&\ctrl{2}&&\\
    & \targX{} && \targ{} & \ctrl{0} & \targ{} &\\
    & \targX{} && \ctrl{-1} & \targ{}& \ctrl{-1} & 
    \end{quantikz}
    \vspace{0.2cm}
    \caption{Fredkin Gate}
    \label{fig:Fred}
\end{figure}

\begin{figure}[ht]
    \centering
    \begin{quantikz}[column sep=1.2ex, row sep=2ex]
    & \ctrl{1}\wire[d][1]{a} &  \\
    & \gate{\text{R}_y(\theta)} &
    \end{quantikz}
    $=$
    \begin{quantikz}[column sep=1.2ex, row sep=2ex]
    &&\ctrl{1} && \ctrl{1} &\\
    &\gate{\text{R}_y(\theta/2)}&\targ{}&\gate{\text{R}_y(-\theta/2)}&\targ{}&
    \end{quantikz}
    \caption{controlled-$\text{R}_y$ Gate}
    \label{fig:CR}
\end{figure}
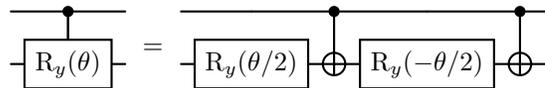

\autoref{tab:Table4} lists the elementary quantum gates used to construct all other required gates. The figures below outline the approaches for decomposing the Toffoli gate (Fig.~\ref{fig:Tof}), Fredkin gate (Fig.~\ref{fig:Fred}), and controlled-$\text{R}_y$ gate (Fig.~\ref{fig:CR}). All of them have a constant depth.

Additionally, a multi-controlled rotation gate can be implemented by combining two multi-controlled Toffoli gates and two single-qubit rotation gates, as shown in Fig.~\ref{fig:CCR}. Similarly, as shown in Fig.~\ref{fig:OR}, a quantum multi-OR-controlled single target X gate can be constructed using a multi-controlled Toffoli gate and the X gates. Consequently, the complexity of both gates depends on the $n$-Toffoli gate, whose size and depth are given in Lemma~\ref{Lm:3} (without ancilla) and Lemma~\ref{Lm:1} (with one ancilla)..

\begin{figure}[H]
    \centering
    \begin{quantikz}[column sep=1.2ex, row sep=2ex]
    & \ctrl{1}\wire[d][1]{a} &  \\
    & \ctrl{0}\wire[d][1]{a} &  \\
    \setwiretype{n} &\vdots &\\
    & \ctrl{2}\wire[u][1]{a} &  \\
    & \ctrl{0}\wire[d][0]{a} &  \\
    & \gate{\text{R}_y(\theta)} &
    \end{quantikz}
    $=$
    \begin{quantikz}[column sep=1.2ex, row sep=2ex]
    && \ctrl{1} & & \ctrl{1} &\\
    && \ctrl{0}\wire[d][1]{q} & & \ctrl{0}\wire[d][1]{q} &\\
    \setwiretype{n} &\vdots &&\vdots &\\
    && \ctrl{2}\wire[u][1]{q} & & \ctrl{2}\wire[u][1]{q} &\\
    && \ctrl{0} & & \ctrl{0} &\\
    &\gate{\text{R}_y(\theta/2)}&\targ{}&\gate{\text{R}_y(-\theta/2)}&\targ{}&
    \end{quantikz}
    \vspace{0.2cm}
    \caption{control-$\text{R}_y$ Gate}
    \label{fig:CCR}
\end{figure}
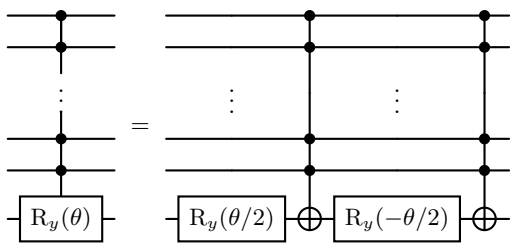
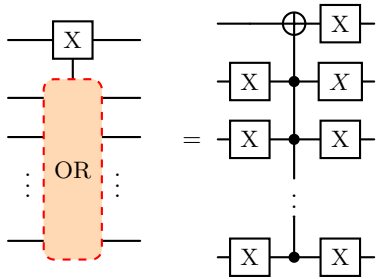
\begin{figure}[H]
    \centering
    \begin{quantikz}[column sep=1.2ex, row sep=2ex]
        & \gate{\text{X}} \vqw{1} &\\
        & \gate[4, style={draw=red, dashed, fill=orange!30, rounded corners}]{\text{OR}} &\\
        &&\\
        \setwiretype{n}\phantom{Pig}\vdots && \vdots\phantom{Dog}\\
        &&
    \end{quantikz}
    $=$
    \begin{quantikz}[column sep=1.2ex, row sep=2ex]
        && \targ{}  & \gate{\text{X}} & \\
        & \gate{\text{X}} & \ctrl{0} & \gate{X} & \\
        & \gate{\text{X}} & \ctrl{-2}\wire[d][1]{q}  & \gate{\text{X}} & \\
        \setwiretype{n} && \vdots  & \\
        & \gate{\text{X}} & \ctrl{0} \wire[u][1]{q} & \gate{\text{X}}&
    \end{quantikz}
    \caption{A quantum multi-OR-controlled single target X gate (hereafter referred to as a quantum OR-controlled X gate), takes multiple control qubits and flips the target qubit if at least one of the control qubits is in the $\ket{1}$ state, i.e. the OR of the control qubits is 1; otherwise, the target qubit remains unchanged.}
    \label{fig:OR}
\end{figure}

\begin{lemma}[Complexity of the $n$-Toffoli gat without ancilla~\cite{Toffoli_2024}]
\label{Lm:3}
With the gates set drawn from $\text{\{U(2),CNOT\}}$, $n$-Toffoli gate can be implemented by a quantum circuit of depth $O(\log^2 n)$ and size $O(n)$ using no ancilla qubit.
\end{lemma}

\begin{lemma}[Complexity of the $n$-Toffoli gate with one ancilla~\cite{Toffoli_2024}]
\label{Lm:1}
With the gates set drawn from $\text{\{U(2),CNOT\}}$, $n$-Toffoli gate can be implemented by a quantum circuit of depth $O(\log n)$ and size $O(n)$ using one clean ancilla qubit.
\end{lemma}

\begin{figure}[H]
    \centering
    \begin{quantikz}[column sep=1.2ex, row sep=2ex]
    &\ctrl{3} & \midstick[4,brackets=none]{=} &\ctrl{1}&\ctrl{2}&\ctrl{3}&\\
    & \targ{} && \targ{} &&&\\
    & \targ{} &&  & \targ{} &&\\
    & \targ{} &&&& \targ{} &
    \end{quantikz}
    \vspace{0.2cm}
    \caption{Quantum fan-out Gate}
    \label{fig:fanout}
\end{figure}
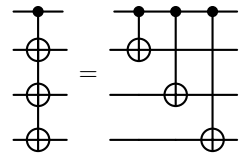

The result of Lemma~\ref{Lm:3} allows us to complete the construction of the multi-controlled rotation gate with only logarithmic depth complexity. However, the quantum fan-out gate shown in Fig.~\ref{fig:fanout} requires a linear depth to implement, which may increase the upper bound of the QSP circuit. As a result, we will discuss $\mathsf{MaF}$ in the next subsection to address this issue.

\subsection{\label{sec:2.2}Quantum circuit with measurement and feedforward}

In contrast to typical quantum circuits, a circuit with measurement and feedforward performs not only measurements at the end of the circuit to extract the result, but also intermediate measurements on selected qubits during the computation. The mid-circuit measurement outcomes are classically processed and used to control subsequent quantum operations. Inspired by the model presented in \cite{shallow_2024}, we define the quantum circuit with measurement and feedforward ($\mathsf{MaF}$) as follows:

\begin{definition}[Quantum circuits with measurement and feedforward]
Let $\mathsf{MaF}(\mathcal{Q}, \mathcal{C}, l)$ be the class of circuits such that
\begin{enumerate}[label=(\arabic*)]
\item every quantum layer implements a quantum circuit $Q\in\mathcal{Q}$;
\item every classical layer implements a classical circuit $C\in\mathcal{C}$;
\item there are $l$ alternating layers of quantum and classical circuits;
\item after every quantum circuit $Q$ a subset of the qubits is measured;
\item the classical circuits receive inputs from the measurement outcomes of previous quantum layers;
\item the classical circuits can control quantum operations in future layers.
\end{enumerate}
\end{definition}
We restrict our attention to a subclass of $\mathsf{MaF}(\mathcal{Q}, \mathcal{C}, l)$, where $\mathcal{Q}=\mathrm{QPoly(n)}$, $\mathcal{C}=\mathrm{NC}^1$, and $l=O(n)$. Throughout this work, we use $\mathsf{MaF}$ to to abbreviate the class $\mathsf{MaF}(\mathrm{QPoly(n)}, \mathrm{NC}^1, O(n))$. Here, $\mathrm{QPoly(n)}$ denotes the class of quantum circuits with polynomial size, and $\mathrm{NC^1}$ denotes the class of classical circuits with logarithmic depth and polynomial size. We refer
to a circuit within this framework as a $\mathsf{MaF}$ circuit.

\begin{figure}[H]
    \centering
    \includegraphics[width=0.9\columnwidth]{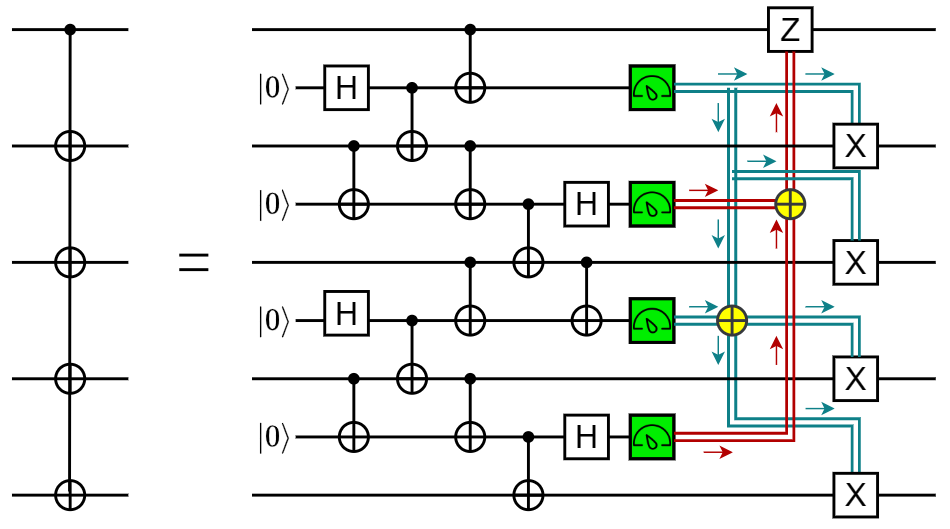}
    \caption{Measurement-based implementation of the fan-out gate}
    \label{fig:LAQCC}
\end{figure}

Among various usages of $\mathsf{MaF}$ circuit, one important use is in reducing circuit depth. We demonstrate how an $n$-qubit fan-out gate can be implemented using a constant depth circuit within $\mathsf{MaF}$ framework, as shown in the following theorem.
\begin{theorem}[Quantum fan-out gate within $\mathsf{MaF}$ framework~\cite{LAQCC_2024}]
\label{th:3}
An $n$-qubit quantum fan-out gate can be implemented by a constant-depth quantum circuit using $n-1$ ancilla qubits, and requires only a single layer of intermediate measurements.
\end{theorem}
The $\mathsf{MaF}$ circuit in Fig.~\ref{fig:LAQCC}  is an example of a $5$-qubit quantum fan-out gate. It adds $4$ ancilla qubits initialized in $\ket{0}$. The first quantum layer in $\mathcal{Q}$ consists of a series of Hadamard and CNOT gates, followed by the mid-circuit measurements. The only classical layer $C$ performs parallel XOR operations based on the measurement outcomes. Finally, the second quantum layer in $\mathcal{Q}$, which includes the final X and Z gates, is controlled by the classical output. 
\vspace{-5pt}
Notably, the fan-out gate requires only a single round of mid-circuit measurement, resulting in $l = 1$ alternating layer of quantum and classical circuits. This structure enables its implementation on an arbitrary number of qubits with constant depth while maintaining linear size complexity, which is crucial to achieve SQSP circuit depth independent of the sparsity $d$.

\subsection{Notation}

In this paper, we adopt the following notation throughout the pseudocode. For a quantum register R $\in \{$A, B, C, D$\}$, the $i$th qubit in the R register is denoted as $\text{R}(i)$. For example, $\text{A}(i)$ refers to the $i$th qubit in the A register. When referring to a consecutive subset of qubits in a register, we use $\text{R}(i,j)$ to denote the set of qubits from index $i$ to $j$ in the R register.

Building upon the notation defined above, we now introduce the expression of single-control quantum gates. A controlled-NOT gate acting on control qubit $c$ and target qubit $t$ is denoted as $\text{CNOT}(c, t)$. Similarly, a controlled-SWAP gate acting on control qubit $c$ and swapping the two target qubits $t_1$ and $t_2$ is written as $\text{CSWAP}(c, t_1, t_2)$. The notation $\text{FANOUT}(c, t)$ represents a fan-out gate, where $c$ is the control qubit and $t$ is a set of target qubits. Both the control and target qubits in these gates are specified using the register notation introduced earlier. For example, $\text{FANOUT}(\text{B}(i), \text{A}(j:k))$ applies X gates to the $j$th through $k$th qubits in the A register, controlled by the $i$th qubit in the B register.

We next define multi-controlled quantum operations. Let CQ denote a classical register that stores the indices of control qubits. Based on this, we use the notation $\text{OR-C-X}(\text{R(CQ)}, t)$ to represent a quantum multi-OR-controlled single-target X gate, and $\text{Par-C-X}(\text{R(CQ)}, t)$ to represent a parity-controlled X gate, where $t$ is the target qubit and the control qubits are those specified by the indices in $\text{CQ}$ within the R register.

We also define the notation $i(j)$ and $q_i(j)$ to refer to the value of the $j$th bit in the binary representation of $i$ and $q_i$, respectively. The bits are indexed from left to right. For instance, if $i=4$ and its binary representation is $100$, then $i(0)=1$, $i(1)=0$, and $i(2)=0$.

\subsection{General quantum state preparation algorithm}

The Grover-Rudolph algorithm~\cite{GRQSP_2002} is a foundational method for general quantum state preparation (GQSP), and serves as the basis for many subsequent developments in this area. We briefly review this algorithm in this subsection. Nevertheless, the GQSP subroutine utilized in our SQSP algorithm differs from the original Grover–Rudolph method. Instead, we adopt a low-depth GQSP algorithm, as described in Theorem~\ref{th:5}.

The General Quantum State Preparation problem aims to create an arbitrary quantum state of the form:
\begin{equation}
    \ket{\psi}=\sum_{i=0}^{2^n-1}\alpha_i\ket{i}=\sum_{i=0}^{2^n-1}\alpha_i\ket{i_0...i_{n-1}},
\end{equation}
where $n$ denotes the number of qubits, $\alpha_i$ denotes the amplitude corresponding to $\ket{i}$, and $i_0...i_{n-1}$ is the binary representation of $i$.

\begin{theorem}[Grover-Rudolph algorithmm~\cite{GRQSP_2002}]
\label{th:4}
With the gate set drawn from $\text{\{U(2),CNOT\}}$, an arbitrary $n$-qubit quantum state can be prepared with a circuit size $O(n2^n)$, depth\footnote{While the circuit depth was not provided in~\cite{GRQSP_2002}, we derive the depth by applying a recent low-depth construction of multi-controlled Toffoli gates.} $O(\log^2 n\cdot2^n)$.
\end{theorem}

To prepare the state $\ket{\psi}$, we need to implement a unitary $U_\psi$ such that $U_\psi\ket{0}=\ket{\psi}$. The strategy is to decompose $U_\psi$ into a series of simpler unitary transformations, $U_\psi = U_{n-1}...U_0$, where each unitary $U_i$ performs the transformation $U_i\ket{\psi_i}=\ket{\psi_{i+1}}$. We define the intermediate states $\ket{\psi_i}$ as follows:
\begin{equation}
    \label{eq:3}
    \ket{\psi_i}=\sum_{k=0}^{2^i-1}x_{i,k}\ket{k}\ket{0}^{\otimes n-i}=\sum_{k=0}^{2^i-1}x_{i,k}\ket{k_0...k_{i-1}}\ket{0}^{\otimes n-i},
\end{equation}
where $k_0 \cdots k_{i-1}$ is the binary representation of $k$, and $x_{i,k}$ is defined recursively by:
\begin{equation}\label{eq:apt}
    x_{i,k}=\sqrt{x_{i+1,2k}^2+x_{i+1,2k+1}^2}
\end{equation}
with $x_{n,k}=\alpha_k$, for all $k$. Clearly, $\ket{\psi_0}$ is the initial state $\ket{0}^{\otimes n}$ of the algorithm, and $\ket{\psi_n}$ corresponds to the desired output state $\ket{\psi}$.

Each unitary $U_i$ can be implemented by applying a series of controlled rotations on the $i$th qubit, such that $U_i$ transforms $\ket{\psi_i}$ into $\ket{\psi_{i+1}}$ as follows:
\begin{equation}
    \label{eq:2}
    \begin{aligned}
    &\quad \; U_i\ket{\psi_i} \\[1ex] 
    &= U_i\sum_{k=0}^{2^i-1}x_{i,k}\ket{k}\ket{0}^{\otimes n-i}\\[1ex] &=\sum_{k=0}^{2^{i+1}-1}x_{i,k}\ket{k}(\cos\left({\frac{\theta_{i,k}}{2}}\right)\ket{0}+\sin\left(\frac{\theta_{i,k}}{2}\right)\ket{1})\ket{0}^{\otimes n-i-1}\\[1ex]
    &=\ket{\psi_{i+1}}.
    \end{aligned}
\end{equation}
Here, the rotation angle $\theta_{i,k}$ is defined as:
\begin{equation}
    \label{eq:1}
    \theta_{i,k} = 2\arccos\left(\frac{x_{i+1,2k}}{x_{i,k}}\right),
\end{equation}

To efficiently store and compute the intermediate amplitudes $x_{i,k}$ and rotation angles $\theta_{i,k}$, we utilize a binary search tree (BST) data structure. The BST is constructed recursively, where the $i$th leaf node is assigned the amplitude $x_{n,i} = \alpha_i$. Each internal node stores the value $x_{i,k}$ computed recursively using Eq.~\eqref{eq:apt}, along with its corresponding rotation angle $\theta_{i,k}$ as defined in Eq.\eqref{eq:1}.  The recursion proceeds from the leaf nodes up to the root, with each internal node's amplitude computed based on its two child nodes. As a result, the root node holds the value $x_{0,0} = 1$ and the height of the BST will necessarily be $n+1$. Each edge from a parent node to its left child is labeled with a $0$, while each edge to its right child is labeled with a $1$. As a result, each path from the root to a leaf encodes a unique basis state in bianry, corresponding to the position of the amplitude $\alpha_i$ in the final quantum state.

The rotation angles $\theta_{i,k}$ stored in the BST serve as the parameters for constructing the quantum circuit. Specifically, the unitary $U_i$ can be further decomposed into a sequence of $2^i$ multi-controlled single-rotation gates of the form $\text{CR}_y(\theta_{i,k})$. Each $\text{CR}_y(\theta_{i,k})$ gate is controlled by the first $i$ qubits, and applies a $y$-axis rotation $\text{R}_y(\theta_{i,k})$ on the $i+1$th qubit when the control qubits are in the basis state $\ket{k}$, for $k \in {0,1,\dots,2^i-1}$. The rotation acts as follows:
\begin{equation}
R_y(\theta_{i,k})\ket{0} = \cos\left(\frac{\theta_{i,k}}{2}\right)\ket{0} + \sin\left(\frac{\theta_{i,k}}{2}\right)\ket{1}.
\end{equation}

Once all $\text{CR}_y(\theta_{i,k})$ operations for $k = 0, \dots, 2^i - 1$ are applied, the unitary $U_i$ completes its task. It adjusts the amplitude of the $i+1$th qubit conditioned on the first $i$ qubits, resulting in the updated state $\ket{\psi_{i+1}}$. This method allows the algorithm to build the desired superposition one qubit at a time, with the correct amplitudes encoded through the controlled rotations. Hence, after executing $U_{n-1}$ on $\ket{\psi_{n-1}}$, the target state $\ket{\psi_n} = \ket{\psi}$ is successfully prepared. The procedure explained above are depicted in Algorithm~\ref{alg:GRA}.

\begin{PRAalgorithm}[Grover-Rudolph algorithm.]{alg:GRA}
\renewcommand{\algorithmicrequire}{\textbf{Input:}}
\renewcommand{\algorithmicensure}{\textbf{Output:}}
    \Require $\ket{0}^n$, $\alpha_k$, for all $k$
\begin{spacing}{1.5}
    \Ensure $\sum\limits_{i=0}^{2^n-1}\alpha_i\ket{i}$
\end{spacing}

\For{$k \gets 0$ to $2^n$-1}
    \State $x_{n,k} \gets \alpha_k$
\EndFor

\For{$i\gets n-1$ to $0$}
    \For{$k \gets$ to $2^i-1$}
        \State $x_{i,k} \gets \sqrt{x_{i+1,2k}^2+x_{i+1,2k+1}^2}$
        \State $\theta_{i,k} \gets 2\arccos{\left(\frac{x_{i+1,2k}}{x_{i,k}}\right)}$
    \EndFor
\EndFor

\For{$i \gets 0$ to $n-1$}
\Comment{Implement $U_i$ in Eq.\eqref{eq:2}}
    \For{$k \gets 0$ to $2^i-1$}
    \Comment{$k_{(10)}=k_0..k_{i-1\ (2)}$}
        \State $\mathrm{CR}_y(\theta_{i,k_0...k_{i-1}})$
    \EndFor
\EndFor
\end{PRAalgorithm}

Here, we present a simple example of a 3-qubit quantum state to illustrate general quantum state preparation:
\begin{equation}
    \begin{aligned}
    &\ket{\psi}=\sqrt{0.05}\ket{000}+\sqrt{0.1}\ket{001}+\sqrt{0.03}\ket{010}+\sqrt{0.17}\ket{011}\\
    &\quad \;\; +\sqrt{0.35}\ket{100}+\sqrt{0.09}\ket{101}+\sqrt{0.18}\ket{110}+\sqrt{0.03}\ket{111}.
    \end{aligned}
\end{equation}

\begin{figure}[ht]
    \centering
    \includegraphics[width=0.9\columnwidth]{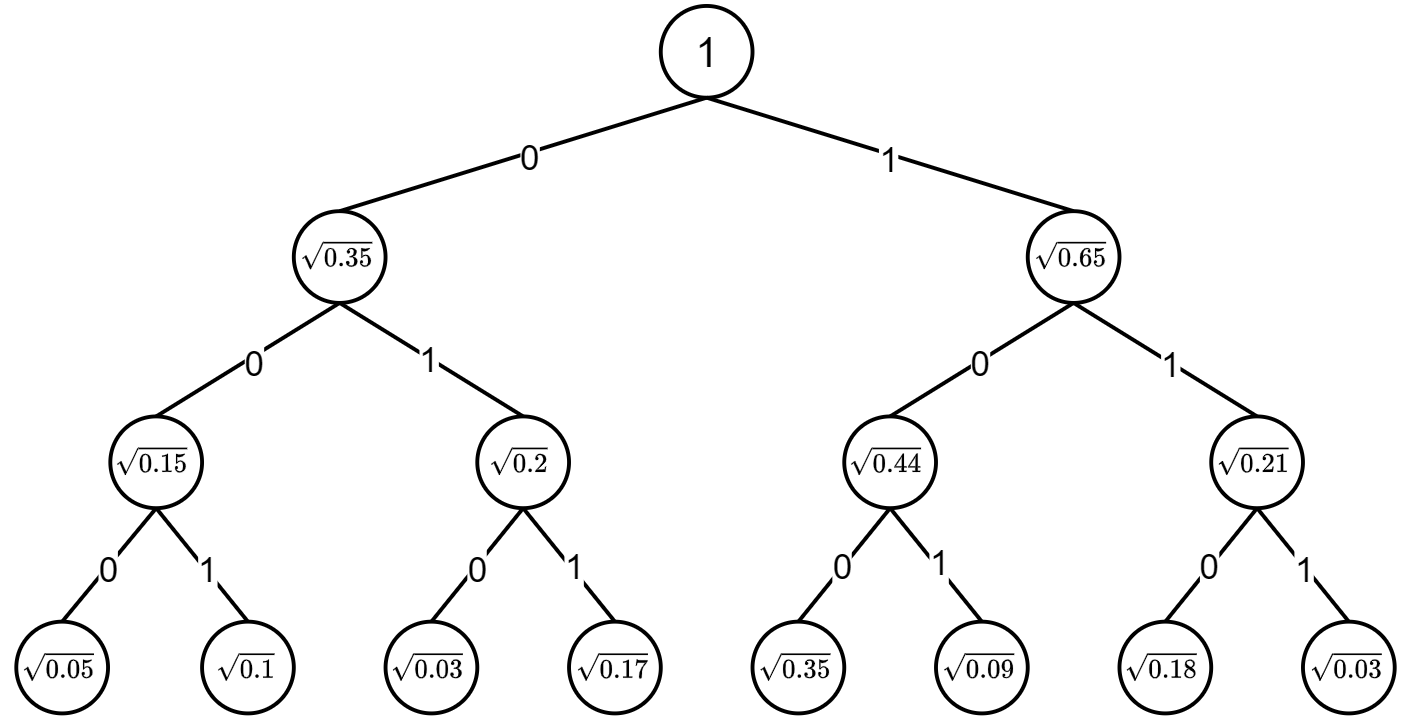}
    \caption{Binary search tree with $x_{i,k}$ for 3-qubit state $\ket{\psi}$}
    \label{fig:BST}
\end{figure}

The first step is to classically compute all the values of $x_{i,k}$ by building the BST corresponding to $\ket{\psi}$, as depicted in Fig.~\ref{fig:BST}.

We can then obtain the intermediate states $\ket{\psi_i}$:
\begin{equation}
    \begin{aligned}
        &\ket{\psi_0}=\ket{000},\\
        &\ket{\psi_1}=(\sqrt{0.35}\ket{0}+\sqrt{0.65}\ket{1})\ket{00},\\
        &\ket{\psi_2}=(\sqrt{0.15}\ket{00}+\sqrt{0.2}\ket{01}+\sqrt{0.44}\ket{10}+\sqrt{0.21}\ket{11})\ket{0},\\
        &\ket{\psi_3}=\ket{\psi}.
    \end{aligned}
\end{equation}
\noindent

The rotation angles $\theta_{i,k}$ can also be computed from the stored amplitudes using Eq.~\eqref{eq:1}.

\begin{figure}[ht]
    \centering
    \includegraphics[width=0.9\columnwidth]{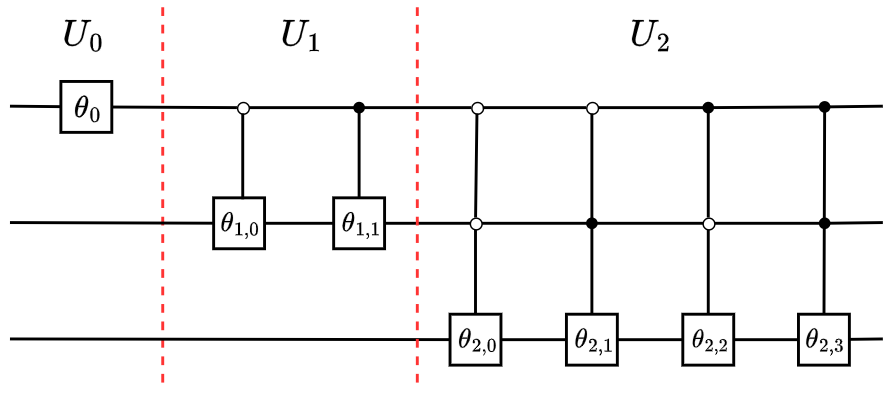}
    \caption{The Grover-Rudolph circuit to prepare the 3-qubit state $\ket{\psi}$}
    \label{fig:GQSP}
\end{figure}

The circuit, as depicted in Fig.~\ref{fig:GQSP}, is composed of $U_0, U_1$ and $U_2$. The function of $U_0$ is to map $\ket{000}$ to $\ket{\psi_1}$. It can be done by rotating the first qubit with angle $\theta_0$. Similar to $U_0$, $U_1$ also applies a rotation on the second qubit, but the rotation angle depends on the first qubit. Since the first qubit has two possible states, $\ket{0}$ and $\ket{1}$, $U_1$ has two different angles corresponding to each case. In the same way, $U_2$ consists of four different rotation gates with controlled gates acting on the first two qubits. The Grover-Rudolph procedure sequentially rotates each qubit with specific angles, eventually leading to the final state $\ket{\psi_3}\equiv\ket{\psi}$.

However, due to the exponential circuit depth required by Grover-Rudolph's algorithm, it cannot be completed quickly. To address this, other studies have proposed using additional ancilla qubits to parallelize the algorithm and reduce the circuit depth.

\begin{theorem}[Low-depth GQSP algorithm~\cite{STQSP_2024}]
\label{th:5}
With the gates set drawn from $\text{\{U(2),CNOT\}}$, an arbitrary $n$-qubit quantum state can be prepared with a circuit depth $O(n)$ and $O(2^n)$ ancilla qubits.
\end{theorem}

\section{Sparse Quantum State Preparation Algorithm}\label{chap:SQSP}
In this section, we present algorithm for preparing any $n$-qubit $d$-sparse quantum state 
\begin{equation}
    \ket{\phi}=\sum_{i=0}^{d-1}\alpha_i\ket{q_i}.
\end{equation}
 using $O(d)$ ancilla qubits. The primary goal is to reduce the circuit depth for SQSP. The algorithm is made up of four steps: GQSP, one-hot encoding, permutation, and garbage elimination, as shown in the Eq.\eqref{eq:SQSP}.

\begin{equation}
    \label{eq:SQSP}
    \begin{aligned}
    &\ket{0}^{\otimes n}\ket{0}^{\otimes 4d}\\
    \xrightarrow{1}\ & \sum^{d-1}_{i=0}\alpha_i\ket{i}\ket{0}^{\otimes n-\lceil\log d\rceil}\ket{0}^{\otimes 4d}\\
    \xrightarrow{2}\ & \sum^{d-1}_{i=0}\alpha_i\ket{i}\ket{0}^{\otimes n-\lceil\log d\rceil}\ket{e_i}\ket{0}^{\otimes 3d}\\
    \xrightarrow{3}\ & \sum^{d-1}_{i=0}\alpha_i\ket{q_i}\ket{e_i}\ket{0}^{\otimes 3d}\\
    \xrightarrow{4}\ & \sum^{d-1}_{i=0}\alpha_i\ket{q_i}\ket{0}^{\otimes 4d}\\    
    =\ & \ket{\phi}\ket{0}^{\otimes 4d}.
    \end{aligned}
\end{equation}

The first step, GQSP, uses the GQSP algorithm of Theorem~\ref{th:5} on the first $\lceil\log d\rceil$ qubits to generate $d$ amplitudes, corresponding to $\alpha_0$ through $\alpha_{d-1}$ of the target state $\ket{\phi}$. In the second step, to enable parallel execution of subsequent procedures, we perform one-hot encoding that maps the $d$ basis states to $d$ ancilla qubits. The third step, permutation, uses the one-hot state as control to permute all amplitudes to their correct positions. Finally, after resetting the ancilla qubits, the algorithm is complete and the target state $\ket{\phi}$ is obtained. In the following, we will walk through the details of 4 steps.

\subsection{General quantum state preparation\label{Step1}}

The first step is to construct a $\lceil\log d\rceil$-qubit dense state $\ket{\phi'}$:
\begin{equation}
    \ket{\phi'} = \sum_{i=0}^{d-1}\alpha_i\ket{i}.
\end{equation}
To reduce the circuit depth, we adopt the GQSP algorithm described in Theorem~\ref{th:5}, which prepares $\ket{\phi'}$ using $O(\log d)$ circuit depth and $O(d)$ ancilla qubits.

\subsection{One-hot encoding\label{Step2}}

For convenience, we refer to the state $\ket{\phi'}\ket{0}^{n-\lceil\log d\rceil}$ as the A register. The $2d$ ancilla qubits are divided into two $d$-qubit registers, referred to as the B register and the C register.

To further reduce the circuit depth with ancilla qubits, we map the first $\lceil\log d\rceil$-qubits of register A, $\ket{i}$, to the B register as one-hot encoding:

\ifreprint
\begin{equation}
    \begin{aligned}
        &\sum^{d-1}_{i=0}\alpha_i
        \underbrace{\ket{i}\ket{0}^{n-\lceil\log d\rceil}}_{A}
        \underbrace{\ket{0}^{\otimes d}}_{B}
        \underbrace{\ket{0}^{\otimes d}}_{C} \\
        \longrightarrow\ & \sum^{d-1}_{i=0}\alpha_i
        \ket{i}\ket{0}^{n-\lceil\log d\rceil}
        \ket{e_i}\ket{0}^{\otimes d},
    \end{aligned}
\end{equation}
\else
\begin{equation}
    \sum^{d-1}_{i=0}\alpha_i
    \underbrace{\ket{i}\ket{0}^{n-\lceil\log d\rceil}}_{A}
    \underbrace{\ket{0}^{\otimes d}}_{B}
    \underbrace{\ket{0}^{\otimes d}}_{C}
    \longrightarrow
    \sum^{d-1}_{i=0}\alpha_i
    \ket{i}\ket{0}^{n-\lceil\log d\rceil}
    \ket{e_i}\ket{0}^{\otimes d},
\end{equation}
\fi
where $\ket{e_i}$ denotes the quantum state in which the $i$th qubit is $\ket{1}$, while all other qubits are $\ket{0}$, known as the one-hot encoding of $i$.

\begin{figure}[ht]
    \centering
    \includegraphics[width=0.9\columnwidth]{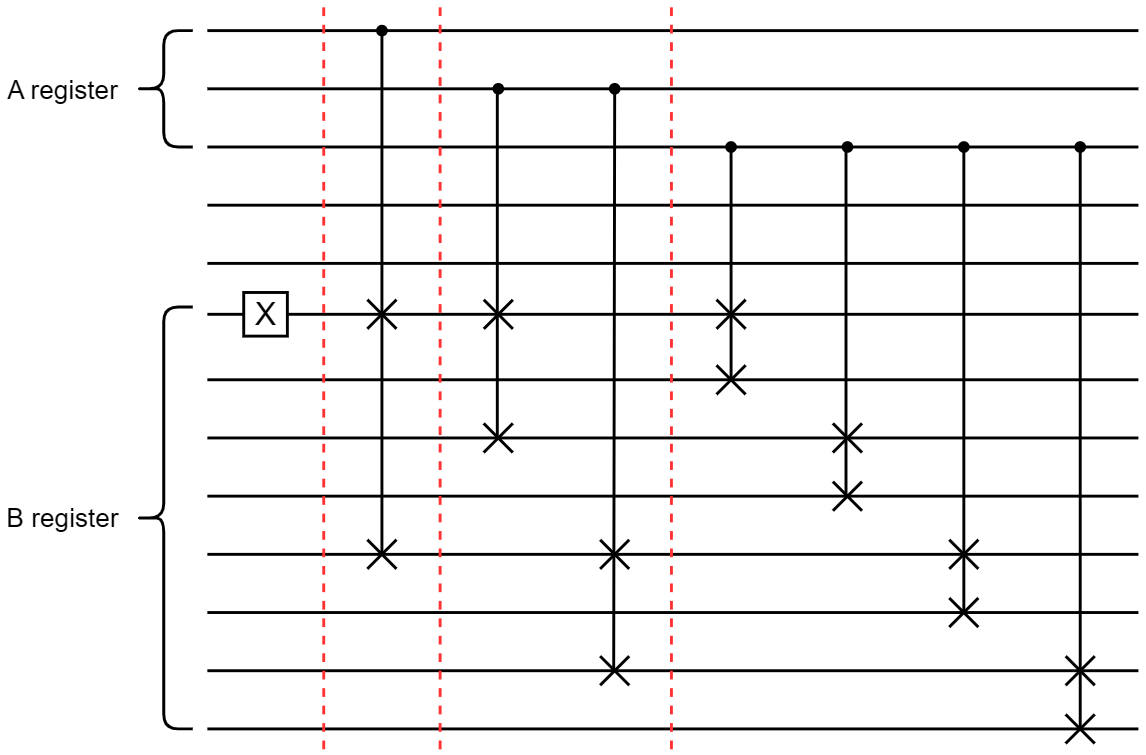}
    \caption{A simple example of one-hot encoding for $n=5, d=8$}
    \label{fig:Step2}
\end{figure}

We can implement this transformation using an X gate and CSWAP gates. As an example, Fig.~\ref{fig:Step2} illustrates a circuit for executing the one-hot encoding when $n=5, d=8$. The procedure begins by applying an X gate to the first qubit of the B register, resulting in the state $\ket{e_0}$. Next, we will use the A register to control a sequence of CSWAP gates that move the $\ket{1}$ to its correct position. For $j=0\sim \lceil\log d\rceil-1$, the $j$th qubit in A register controls $2^j$ CSWAP gates, with their first target qubits uniquely corresponding to all possible $\ket{1}$ positions. These possibe positions are, in fact, the $2^j$ target qubits of the CSWAP gates controlled by the $j-1$th qubit in A register. Each of CSWAP gates controlled by $j$th qubit can be discussed in two cases. If the $j$th qubit is $\ket{0}$, it means that this qubit does not alter the value of $i$, and thus, the state in the B register does not need to change. If the control qubit is $\ket{1}$, it indicates the addition of a value $2^{\lceil\log d\rceil-j-1}$, so a CSWAP must be applied to move $\ket{1}$ forward by $2^{\lceil\log d\rceil-j-1}$ positions. Consequently, the second target qubits of all the CSWAP gates are positioned $2^{\lceil\log d\rceil-j-1}$th qubits after the first target qubits.

\begin{figure}[h]
    \centering
    \includegraphics[width=\columnwidth]{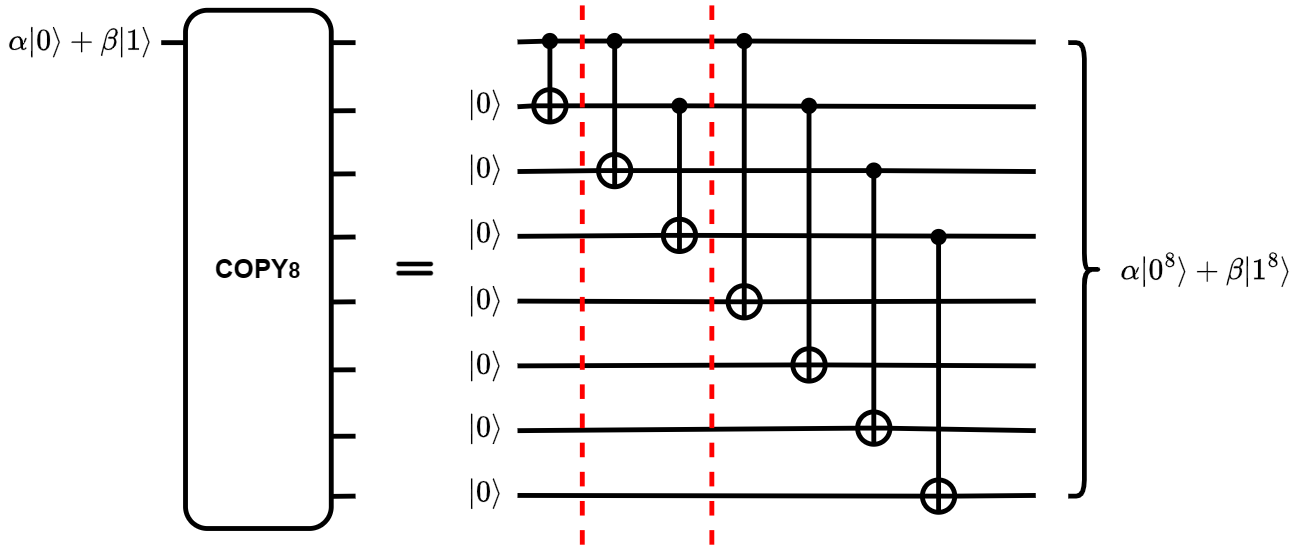}
    \caption{An example of a circuit that performs single qubit copying}
    \label{fig:copy}
    \vspace{-10pt}
\end{figure}

There are $2^j$ swap gates controlled by the $j$th qubit for $j=0\sim \lceil\log d\rceil-1$, so both the circuit size and circuit depth are $O(d)$. In Algorithm~\ref{alg:ISE}, we provide the pseudocode for encoding one-hot state.\\
\begin{PRAalgorithm}[One-hot Encoding.]{alg:ISE}
\renewcommand{\algorithmicrequire}{\textbf{Input:}}
\renewcommand{\algorithmicensure}{\textbf{Output:}}
    \Require $\sum^{d-1}_{i=0}\alpha_i\underbrace{\ket{i}\ket{0}^{\otimes n-\lceil\log d\rceil}}_{A}\underbrace{\ket{0}^{\otimes d}}_{B}$
\begin{spacing}{1.5}
    \Ensure $\sum^{d-1}_{i=0}\alpha_i\ket{i}\ket{0}^{\otimes n-\lceil\log d\rceil}\ket{e_i}$
\end{spacing}

\State Flip the first qubit of the B register.
\For{$j \gets 0$ to $\lceil\log d\rceil-1$}
    \Comment{Designate the $j$th qubit as the control qubit.}
    \For{$i \gets 0$ to $2^j-1$}
        \State CSWAP(A($j$), B($i\cdot2^{\lceil\log d\rceil-j}$), B($i\cdot2^{\lceil\log d\rceil-j}+2^{\lceil\log d\rceil-j-1}$))
    \EndFor
\EndFor
\end{PRAalgorithm}
To reduce the circuit depth, we allow all the CSWAP gates controlled on the same qubit to execute in parallel [The for loop in lines 3 - 5 in Algorithm~\ref{alg:ISE}]. This is achieved by using only CNOT gates to copy enough control qubits into the C register, enabling the parallel execution of all controlled gates on the C register. The copy circuit is illustrated in Fig.~\ref{fig:copy}. In each layer of the circuit, CNOT gates parallel copy the state to fresh ancilla qubits, doubling the number of copies with constant depth. The process of copying a single qubit $k$ times requires only $O(\log k)$ circuit depth.

\begin{PRAalgorithm}[One-hot Encoding with $O(d)$ ancilla qubits.]{alg:C}
\renewcommand{\algorithmicrequire}{\textbf{Input:}}
\renewcommand{\algorithmicensure}{\textbf{Output:}}

    \Require $\sum^{d-1}_{i=0}\alpha_i\underbrace{\ket{i}\ket{0}^{\otimes n-\lceil\log d\rceil}}_{A}\underbrace{\ket{0}^{\otimes d}}_{B}\underbrace{\ket{0}^{\otimes d}}_{C}$
\begin{spacing}{1.5}
    \Ensure $\sum^{d-1}_{i=0}\alpha_i\ket{i}\ket{0}^{\otimes n-\lceil\log d\rceil}\ket{e_i}\ket{0}^{\otimes d}$
\end{spacing}

\State Flip the first qubit of the B register.
\For{$j \gets 0$ to $\lceil\log d\rceil-1$}
    \State CNOT(A($j$), C($0$))
    \For{$k \gets 0$ to $j-1$}
        \For{$l \gets 0$ to $2^k-1$}
            \State CNOT(C($l$), C($l+2^k$))
        \EndFor
    \EndFor
    \For{$i \gets 0$ to $2^j-1$}
        \State CSWAP(C($i$), B($i\cdot2^{\lceil\log d\rceil-j}$), B($i\cdot2^{\lceil\log d\rceil-j}+2^{\lceil\log d\rceil-j-1}$))
    \EndFor
    \For{$k \gets j-1$ to $0$}
        \For{$l \gets 2^k-1$ to $0$}
            \State CNOT(C($l$), C($l+2^k$))
        \EndFor
    \EndFor
    \State CNOT(A($j$), C($0$))
\EndFor
\end{PRAalgorithm}
Originally, the $2^j$ CSWAP gates controlled by the $j$th qubit require a circuit depth of $O(2^j)$. By copying the control qubit to the C register, the CSWAP gates can be controlled by C register in parallel, reducing their circuit depth to a constant. The copying process has a circuit depth $O(j)$. Summing over all $j$ from $0$ to $\lceil\log d\rceil-1$, the total circuit depth of one-hot encoding is $O(\lceil\log d\rceil^2)$. The pseudocode is described in Algorithm~\ref{alg:C}.

\begin{figure}[ht]
    \centering
    \includegraphics[width=0.9\columnwidth]{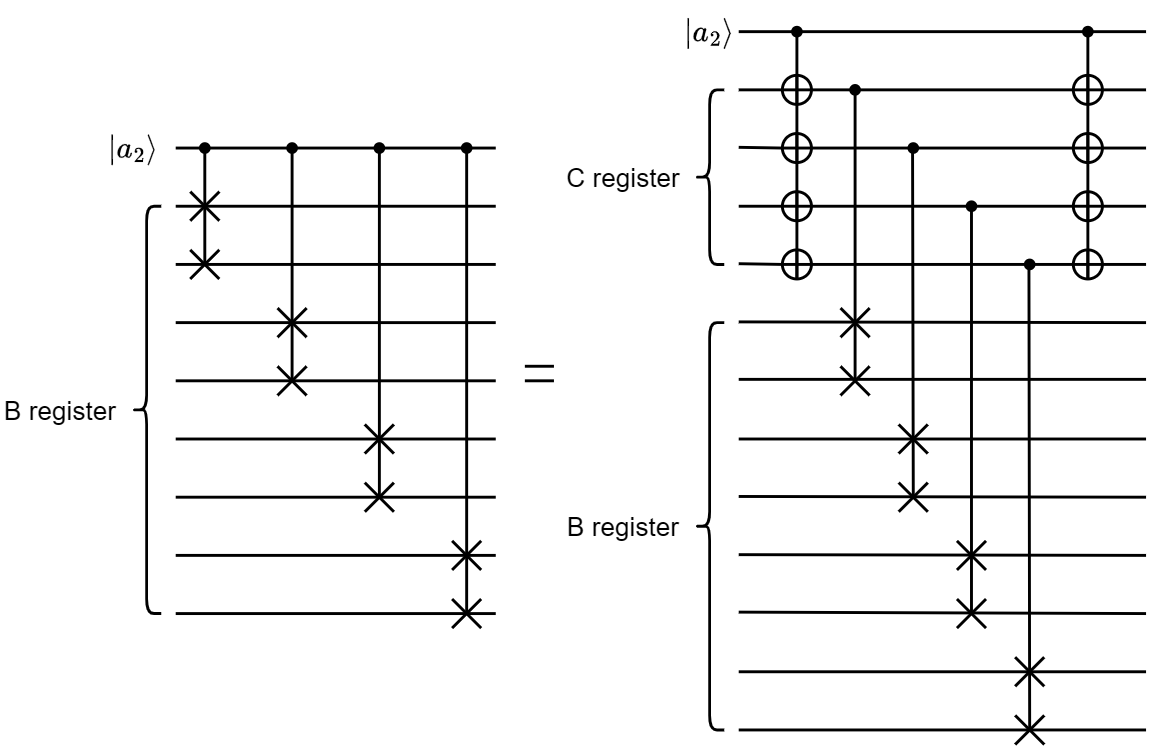}
    \caption{An example of a circuit that performs CSWAP gates in parallel}
    \label{fig:Parallel}
\end{figure}

When the $\mathsf{MaF}$ circuit is allowed, the copy operation can be implemented more efficiently using constant-depth fan-out gates, as depicted in Fig.~\ref{fig:LAQCC}. Since there are at most $\frac{d}{2}$ SWAP gates controlled by the same qubit, the fan-out gate requires at most $\frac{d}{2}$ target qubits and $\frac{d}{2}$ ancilla qubits. Therefore, the C register provides sufficient qubits to perform the copy operation. As a result, Step~\hyperref[Step2]{2} can be completed with circuit depth $O(\log d)$, while maintaining a circuit size of $O(d)$. We give an example of Parallel CSWAP gates in Fig.~\ref{fig:Parallel} and pseudocode for encoding the one-hot state with $\mathsf{MaF}$ circuit in Algorithm~\ref{alg:PISE}.

\begin{PRAalgorithm}[One-hot Encoding with $\mathsf{MaF}$.]{alg:PISE}
\renewcommand{\algorithmicrequire}{\textbf{Input:}}
\renewcommand{\algorithmicensure}{\textbf{Output:}}
    \Require $\sum^{d-1}_{i=0}\alpha_i\underbrace{\ket{i}\ket{0}^{\otimes n-\lceil\log d}\rceil}_{A}\underbrace{\ket{0}^{\otimes d}}_{B}\underbrace{\ket{0}^{\otimes d}}_{C}$
\begin{spacing}{1.5}
    \Ensure $\sum^{d-1}_{i=0}\alpha_i\ket{i}\ket{0}^{\otimes n-\lceil\log d\rceil}\ket{e_i}^{\otimes d}\ket{0}^{\otimes d}$
\end{spacing}

\State Flip the first qubit of the B register.
\For{$j \gets 0$ to $\lceil\log d\rceil-1$}
    \State FANOUT(A($j$), C($0:2^j-1$))
    \For{$i \gets 0$ to $2^j-1$}
        \State CSWAP(C($j$), B($i\cdot2^{\lceil\log d\rceil-j}$), B($i\cdot2^{\lceil\log d\rceil-j}+2^{\lceil\log d\rceil-j-1}$))
    \EndFor
    \State FANOUT(A($j$), C($0:2^j-1$))
\EndFor
\end{PRAalgorithm}

\subsection{Permutation\label{Step3}}
In Step~\hyperref[Step3]{3}, we map the $n$ qubits in the A register to their correct positions, as shown in the transformation below:

\begin{equation}
    \sum^{d-1}_{i=0}\alpha_i\underbrace{\ket{i}\ket{0}^{\otimes n-\lceil\log d\rceil}}_{A}\underbrace{\ket{e_i}}_{B}\underbrace{\ket{0}^{\otimes d}}_{C} \longrightarrow \sum^{d-1}_{i=0}\alpha_i\ket{q_i}\ket{e_i}\ket{0}^{\otimes d}.
\end{equation}

By controlling on the one-hot state, we can identify and flip the properly qubits in the A register one by one. Suppose we are mapping the $j$th qubit in the A register. If the $j$th qubit of $\ket{i}\ket{0}^{\otimes n-\lceil\log d\rceil}$ differs from that of $\ket{q_i}$, a CNOT gate is applied with the $i$th qubit of the B register as the control and the $j$th qubit of the A register as the target. As long as the control qubit is in the state $\ket{1}$, it indicates that the corresponding $j$th qubit in the A register is incorrect and will be flipped by the target gate to reach the expected value. 

Processing the $j$th qubit of $\ket{e_i}$ require at most $d$ CNOT gates with disjoint control but the same target qubit, which costs a high circuit depth of $O(d)$. To minimize the circuit depth, we adopt a quantum OR-controlled X gate to perform the flip operation in parallel, as shown in Fig.~\ref{fig:OR}. The target of the quantum OR-controlled X gate is the $j$th qubit of the A register. For each basis state $\ket{q_i}$, if its $j$th qubit differs from that of the A register, then the $i$th qubit of the B register is included as a control in the  OR-controlled X gate. As long as at least one control qubit is in the $\ket{1}$ state, the target qubit will be flipped.

Since there are at most $d$ control qubits, flipping any qubit in the A register requires a circuit of size $O(d)$ and depth $O(\log d)$. Therefore, mapping all $n$ qubits to their correct positions costs circuit size $O(dn)$ and depth $O(n\log d)$. The pseudocode is described in Algorithm~\ref{alg:PMT1}. 

\begin{PRAalgorithm}[Permutation.]{alg:PMT1}
\renewcommand{\algorithmicrequire}{\textbf{Input:}}
\renewcommand{\algorithmicensure}{\textbf{Output:}}
    \Require $\sum^{d-1}_{i=0}\alpha_i\underbrace{\ket{i}\ket{0}^{\otimes n-\lceil\log d\rceil}}_{A}\underbrace{\ket{e_i}}_{B}\underbrace{\ket{0}^{\otimes d}}_{C}$
\begin{spacing}{1.5}
    \Ensure $\sum^{d-1}_{i=0}\alpha_i\ket{q_i}\ket{e_i}\ket{0}^{\otimes d}$
\end{spacing}

\For{$j \gets 0$ to $n-1$}
    \State CQ $\gets \varnothing$
    \Comment{CQ is a classical variable storing a set of indexes.}
    \For{$i \gets 0$ to $d-1$}
        \If {$i(j) \neq q_i(j)$}
            \State CQ $\gets$ CQ $\bigcup \{i\}$
        \EndIf
    \EndFor
    \State OR-C-X(B(CQ), A($j$))
\EndFor
\end{PRAalgorithm}
Due to the property established in Lemma~\ref{Lm:2}, the parity-controlled X gate can be used in place of the quantum OR-controlled X gate to achieve the same effect.

For example, consider the transformation of 5-qubit 8-sparsity quantum state:

\ifreprint
\begin{equation}
    \begin{aligned}
        &\ket{\psi}=(\alpha_0\ket{000}+\alpha_1\ket{001}+\alpha_2\ket{010}+\alpha_3\ket{011}+\\ 
        &\qquad \ \alpha_4\ket{100}+\alpha_5\ket{101}+\alpha_6\ket{110}+\alpha_7\ket{111})\ket{00}\\
        &\longrightarrow\\ &\alpha_0\ket{01001}+\alpha_1\ket{01110}+\alpha_2\ket{10001}+\alpha_3\ket{10010}\\
        &+\alpha_4\ket{10111}+\alpha_5\ket{11010}+\alpha_6\ket{11101}+\alpha_7\ket{11110}.
    \end{aligned}
\end{equation}  
\else
\begin{equation}
    \begin{aligned}
        &\ket{\psi}=(\alpha_0\ket{000}+\alpha_1\ket{001}+\alpha_2\ket{010}+\alpha_3\ket{011}+\alpha_4\ket{100}+\alpha_5\ket{101}+\alpha_6\ket{110}+\alpha_7\ket{111})\ket{00}\\
        &\longrightarrow\\        &\alpha_0\ket{01001}+\alpha_1\ket{01110}+\alpha_2\ket{10001}+\alpha_3\ket{10010}+\alpha_4\ket{10111}+\alpha_5\ket{11010}+\alpha_6\ket{11101}+\alpha_7\ket{11110}.
    \end{aligned}
\end{equation}
\fi

\begin{figure}[ht]
    \centering
    \includegraphics[width=0.7\columnwidth]{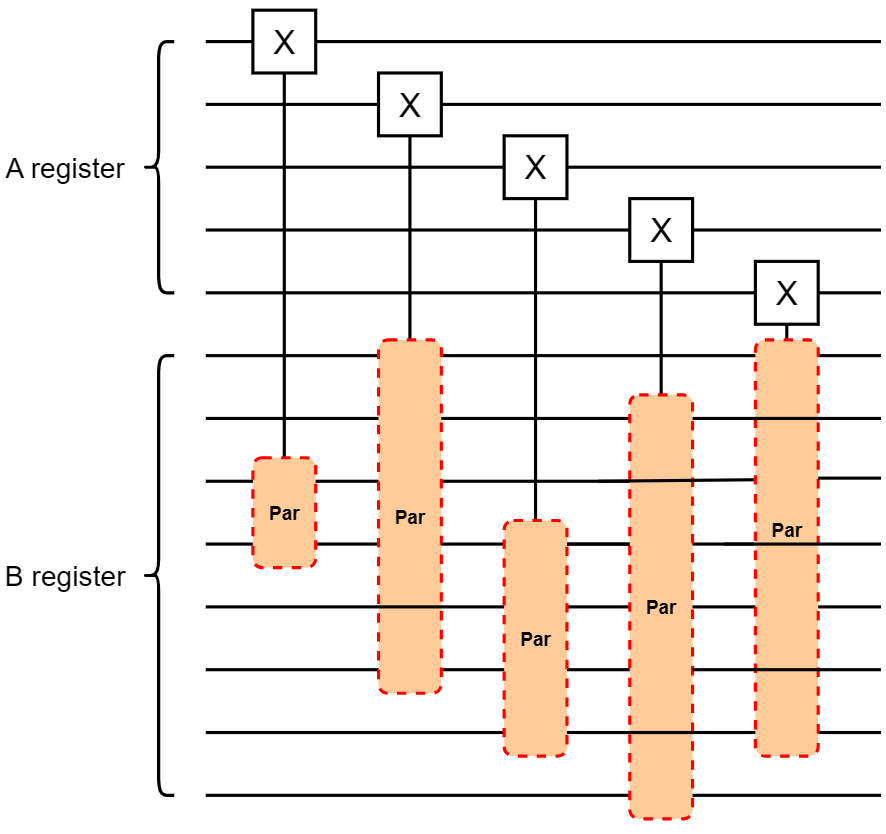}
    \caption{An example circuit for implementing permutation}
    \label{fig:Permutation}
\end{figure}

We can sequentially apply five parity-controlled X gates to the five qubits in the A register, each controlled by the one-hot state, as Fig.~\ref{fig:Permutation}.

\begin{figure}[ht]
    \centering
    \includegraphics[width=\columnwidth]{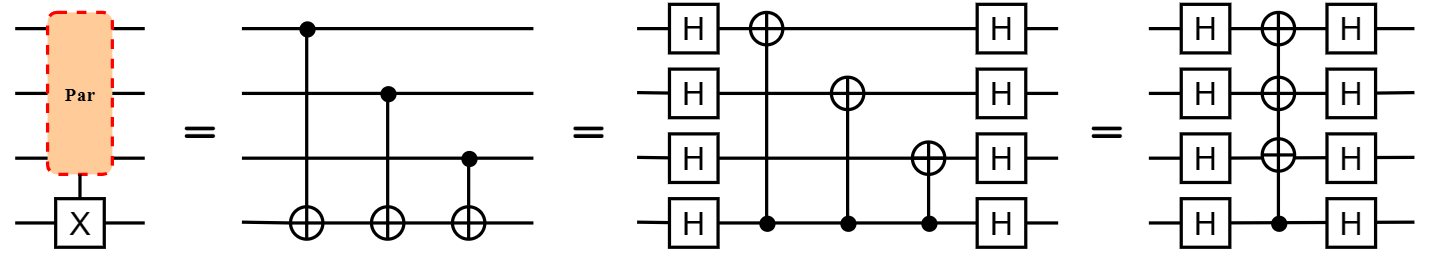}
    \caption{The parity-controlled X gate can be constructed from a fan-out gate and Hadamard gates.}
    \label{fig:Parity}
\end{figure}

As shown in \cite{QC_1999} , parity-controlled X gates and fan-out gates are related by conjugation with a layer of Hadamard gates. This directly leads to the following corollary:

\begin{corollary}[Quantum parity-controlled X gate with $\mathsf{MaF}$]
The $n$-qubit parity-controlled X gate can be implemented by a constant-depth quantum circuit using $n$ ancilla qubits. This implementation requires the Hadamard gates and a constant-depth fan-out gate, which is realized using a circuit with measurement and feedforward($\mathsf{MaF}$), as described in Theorem~\ref{th:3}. Figure.~\ref{fig:Parity} provides a simple example of this construction.
\end{corollary}

Because the parity-controlled X gate involves at most $d$ control qubits, implementing the fan-out gate with $\mathsf{MaF}$ requires, in the worst case, $d$ ancilla qubits. These ancilla qubits can be provided by the C register.

Since Step~\hyperref[Step3]{3} involves $n$ parity-controlled X gates, each with constant depth, the overall depth complexity of Step~\hyperref[Step3]{3} is $O(n)$. Furthermore, as the fan-out gate has a complexity of linear size, the permutation circuit complexity is $O(dn)$. We document the pseudocode of the permutation circuit implementation in Algorithm~\ref{alg:PMT}.

\begin{PRAalgorithm}[Permutation with $\mathsf{MaF}$.]{alg:PMT}
\renewcommand{\algorithmicrequire}{\textbf{Input:}}
\renewcommand{\algorithmicensure}{\textbf{Output:}}
    \Require $\sum^{d-1}_{i=0}\alpha_i\underbrace{\ket{i}\ket{0}^{\otimes n-\log d}}_{A}\underbrace{\ket{e_i}}_{B}\underbrace{\ket{0}^{\otimes d}}_{C}$
\begin{spacing}{1.5}
    \Ensure $\sum^{d-1}_{i=0}\alpha_i\ket{q_i}\ket{e_i}\ket{0}^{\otimes d}$
\end{spacing}

\For{$j \gets 0$ to $n-1$}
    \State CQ $\gets \varnothing$
    \For{$i \gets 0$ to $d-1$}
        \If{$i(j) \neq q_i(j)$}
            \State CQ $\gets$ CQ $\bigcup \{i\}$
        \EndIf
    \EndFor
    \State Par-C-X(B(CQ), A($j$))
\EndFor
\end{PRAalgorithm}

\subsection{Garbage elimination\label{Step4}}

In this step, we use four registers: the A register and the B register as before, the C register for $2d$ qubits, and the D register for $d$ qubits. This step aims to eliminate the one-hot encoding $\ket{e_i}$ in the B register:

\begin{equation}
    \label{eq:4}
    \sum^{d-1}_{i=0}\alpha_i\underbrace{\ket{q_i}}_{A}\underbrace{\ket{e_i}}_{B}\underbrace{\ket{0}^{\otimes 2d}}_{C}\underbrace{\ket{0}^{\otimes d}}_{D} \longrightarrow \sum^{d-1}_{i=0}\alpha_i\ket{q_i}\ket{0}^{\otimes d}\ket{0}^{\otimes 2d}\ket{0}^{\otimes d}.
\end{equation}

Before designing the circuit, we first construct a binary search tree (BST) representing the states $\ket{q_i}$ for all $i$. Each path from the root to a leaf node corresponds to a specific $\ket{q_i}$. The edge at the $j$th layer denotes the value of the $j$th qubit of $\ket{q_i}$: an edge leading to the left child represents $\ket{0}$, and an edge leading to the right child represents $\ket{1}$. A branch node is defined as a node with two child nodes. The $k$th branch node refers to the $k$th branch node in the BST, indexed by traversing from the root to the leaves, and from left to right within each layer, where $k = 1, 2, \dots, d - 1$. The BST has height $n$, $d$ leaf nodes, and $d - 1$ branch nodes in total. It follows that each layer contains at most $\frac{d}{2}$ branch nodes.

We realize the transformation defined in Eq.\eqref{eq:4} through the following steps:

\begin{equation}
    \label{eq:5}
    \begin{aligned}
    &\sum^{d-1}_{i=0}\alpha_i\ket{q_i}\ket{e_i}\ket{0}^{\otimes 2d}\ket{0}^{\otimes d}\\
    \xrightarrow{4.1}\ & \sum^{d-1}_{i=0}\alpha_i\ket{q_i}\ket{e_i}\bigotimes_{k=1}^{d-1}\ket{f(q_i,k)}\ket{0}^{\otimes d}\\
    \xrightarrow{4.2}\ & \sum^{d-1}_{i=0}\alpha_i\ket{q_i}\ket{0}^{\otimes d}\bigotimes_{k=1}^{d-1}\ket{f(q_i,k)}\ket{0}^{\otimes d}\\
    \xrightarrow{4.3}\ & \sum^{d-1}_{i=0}\alpha_i\ket{q_i}\ket{0}^{\otimes d}\ket{0}^{\otimes 2d}\ket{0}^{\otimes d}\\
    =\ & \ket{\phi}.
    \end{aligned}
\end{equation}

We define the 2-qubit state $\ket{f(q_i,k)}$ in~\eqref{eq:5} based on the $k$th branch node in the BST, such that

\ifreprint
\begin{equation}
\ket{f(q_i,k)} =
\begin{cases}
  \ket{10}, & 
  \parbox[t]{0.6\linewidth}{
    if $q_i$ takes the left branch at the $k$th branch node.}\\[1ex]
  \ket{01}, & 
  \parbox[t]{0.6\linewidth}{
    if $q_i$ takes the right branch at the $k$th branch node.}\\
  \ket{00}, &
  \parbox[t]{0.6\linewidth}{
    if the path of $q_i$ does not pass through the $k$th branch node.
  }
\end{cases}
\end{equation}
\else
\begin{equation}
    \ket{f(q_i,k)}=\begin{cases}
    \ket{10}, & \text{if $q_i$ takes the left branch at the $k$th branch node.}\\
    \ket{01}, & \text{if $q_i$ takes the right branch at the $k$th branch node.}\\
    \ket{00},  & \text{if the path of $q_i$ does not pass through the $k$th branch node.}
  \end{cases}
\end{equation}
\fi

In Step~\hyperref[st4_1]{4.1}, the BST is recorded in the C register. This allows Step~\hyperref[st4_2]{4.2} to eliminate the garbage state in the B register by controlling with the C register. Finally, the C register is reset using the same procedure as in Step~\hyperref[st4_1]{4.1}. We now show the implementation of the process step by step and analyze its circuit complexity.

\stepsection{BST encoding}\label{st4_1}
\indent
In the first substep, for each basis state $\ket{q_i}$ in the superposition, the $k$th branch node of the BST is recorded in the 2-qubit state $\ket{f(q_i,k)}$ in the C register. This process is performed in parallel over all $i$, while the records for each $k$ are generated sequentially in the order of branch node indices. Specifically, if $\ket{q_i}$ passes through the left (right) subtree of the $k$th branch node, we set $\ket{f(q_i,k)} = \ket{10}$ ($\ket{01}$). If $\ket{q_i}$ does not pass through the $j$th branch node, then we set $\ket{f(q_i,j)} = \ket{00}$ and no action is needed. Since there are $d-1$ branch nodes and each requires 2 qubits to encode its status, the C register contains sufficient ancilla qubits to store the traversal information for all paths.

To begin recording the information of the first branch node, we start from the $0$th layer of the BST and search layer by layer until we reach the first branch node. Suppose that it is located at the $j$th layer. We then record this branch node in the first two qubits of the C register, denoted by $\ket{f(q_i, 1)}$. This is done by applying an X gate to the first qubit and then a CSWAP gate controlled by the $j$th qubit of the A register. The possible values of $\ket{f(q_i, 1)}$ are $\ket{10}$ or $\ket{01}$: $\ket{10}$ means that the path of $q_i$ in the BST passes through the left subtree of the first branch node, while $\ket{01}$ means that it passes through the right subtree.

Next, we proceed to record the branch nodes in order, traversing from the root to the leaves and from left to right at each layer. Suppose we are currently recording the $k$th branch node B at the $jth$ layer, and let $\ket{\text{B}}$ denote the 2-qubit state that records B. We define the parent branch node of B, denoted by $\text{B}'$, as the nearest branch node along the path from B to the root of the BST. Since branch nodes are recorded in traversal order, we can ensure that $\text{B}'$ has already been recorded before B. To begin recording B, we first apply a CNOT gate, where the target is the first qubit of $\ket{\text{B}}$, and the control is one of the two qubits that encode $\text{B}'$. There are two cases to consider:

\begin{enumerate}[label=(\arabic*)]
    \item If the 2-qubit state recording $\text{B}'$ is $\ket{00}$, it means that $\text{B}'$ is not on the path of $q_i$. Therefore, $B$ is not on the path of $q_i$ either.
    \item If the 2-qubit state recording $\text{B}'$ is $\ket{10}$ or $\ket{01}$, then the path of $q_i$ traverses the left or right subtree of $\text{B}'$, respectively.
\end{enumerate}
Accordingly, if B lies in the left subtree of $\text{B}'$, we use the first qubit of $\text{B}'$ as the control; otherwise, we use the second qubit. In this way, the target qubit is flipped if and only if the path of $q_i$ passes through B, resulting in $\ket{\text{B}} = \ket{10}$. Otherwise, the state remains $\ket{00}$, indicating that $q_i$ does not pass through B.

Subsequently, similar to the process of recording the first branch, we apply a CSWAP gate to the two qubits of $\ket{\text{B}}$, controlled by the $j$th qubit of the A register. If the target qubits are in the state $\ket{00}$, this indicates that B is not on the path of $q_i$, and the CSWAP gate has no effect on $\ket{\text{B}}$. On the other hand, if the target qubits are in the state $\ket{10}$, the effect of the CSWAP depends on the value of the control qubit: if the control qubit is $\ket{0}$, the target qubits remain unchanged, and the state $\ket{\text{B}}$ stays as $\ket{10}$, signifying that $q_i$ takes the left branch at B; if the control qubit is $\ket{1}$, the target qubits are swapped, resulting in the state $\ket{01}$, which indicates that $q_i$ takes the right branch at B. Finally, this completes the recording of the $k$th branch node, where $\ket{\text{B}} = \ket{f(q_i,k)}$, as described in Eq.~\eqref{eq:5}.

\begin{figure}[H]
    \centering
    \includegraphics[width=0.9\columnwidth]{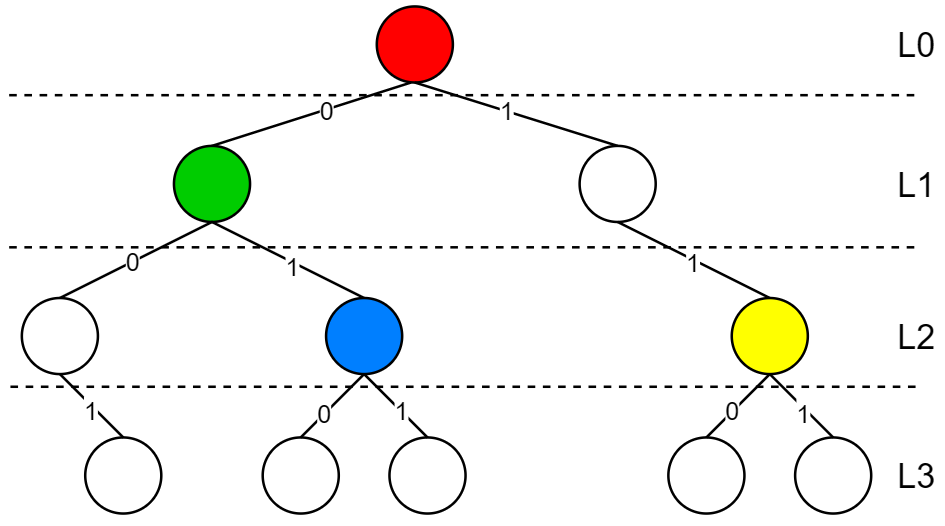}
    \caption{Binary search tree for $\ket{\psi}$. The colored nodes represent the branch nodes.}
    \label{fig:GE_BST}
\end{figure}

\begin{figure}[h!]
    \centering
    \includegraphics[width=0.9\columnwidth]{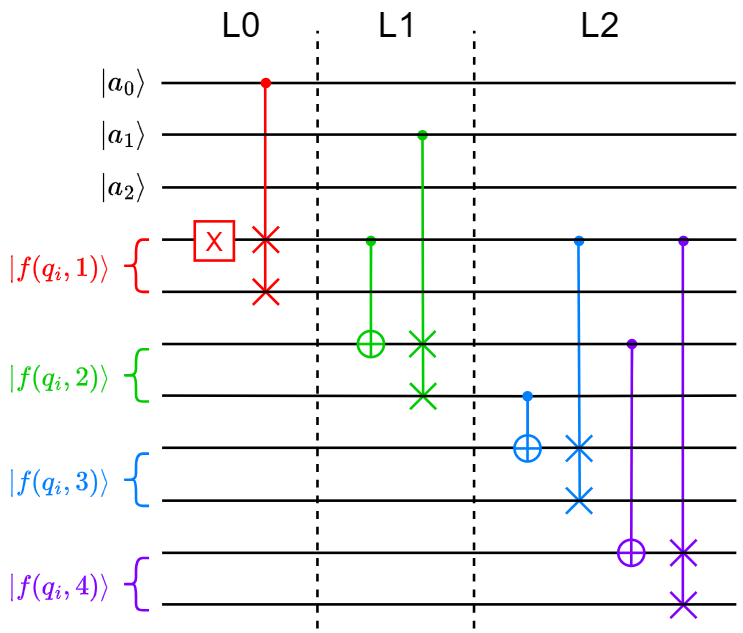}
    \caption{Circuit for recording the binary search tree (BST) encoding without $\mathsf{MaF}$.}
    \label{fig:GE_circuit}
\end{figure}

Here, we present a simple example of BST encoding for a 3-qubit 5-sparsity quantum state:
\begin{equation}
    \begin{aligned}
    &\ket{\psi}=\alpha_0\ket{001}+\alpha_1\ket{010}+\alpha_2\ket{011}+\alpha_3\ket{110}+\alpha_4\ket{110}.
    \end{aligned}
\end{equation}
First, we need to build the BST corresponding to $\ket{\psi}$, as depicted in Fig.~\ref{fig:GE_BST}.
We then record the branch nodes layer by layer, proceeding from the top to the bottom of the BST, and from left to right within each layer. As depicted in Fig.~\ref{fig:GE_circuit}, the circuit consists of three layers. The first layer records the first branch node using an X gate followed by a CSWAP gate. In the second layer, the second branch node is recorded based on the information from its parent—the root node—which has already been recorded. Finally, the third layer records all the branch nodes at the third level of the BST in sequence, using the same method.

Consider the process of recording all $b(j)$ branch nodes located at the $j$th layer of the BST. All the CNOT gates can be executed in parallel since both their control and target qubits are mutually disjoint. However, although the CSWAP gates also act on distinct target qubits, they are all controlled by the same $j$th qubit of the A register. As a result, these CSWAP gates cannot be executed in parallel, which increases the circuit depth.

To reduce the circuit depth, we copy the control qubit into $b(j)$ ancilla qubits in the D register, enabling all the CSWAP gates to be applied simultaneously within the same layer. The resulting circuit depth thus depends on the depth of the copy operation. By using the copy circuit shown in Fig.~\ref{fig:copy}, all branch nodes at the $j$th layer can be recorded in parallel with a circuit size of $O(b(j))$ and a circuit depth of $O(\log b(j))$. If $\mathsf{MaF}$ circuits are allowed, the copy operation can be performed even more efficiently using constant-depth fan-out gates, as illustrated in Fig.~\ref{fig:LAQCC}. Consequently, the recording of all $b(j)$ branch nodes at the $j$th layer can be accomplished in constant circuit depth.

Since the number of branch nodes in any given layer is at most $\frac{d}{2}$, we have $b(j) = O(d)$ for each $j$. In summary, recording the entire BST requires a circuit depth of $O(n \log d)$ without $\mathsf{MaF}$, and $O(n)$ when using the $\mathsf{MaF}$ framework. In both cases, the total circuit size remains $O(d)$. The pseudocode for the full process under the $\mathsf{MaF}$ framework is presented in Algorithm~\ref{alg:RBST}. Here, the notation b(j) denotes the number of branch nodes in the $j$th layer of the BST, and PB(k) refers to the index of the parent branch node of the $k$th branch node in the BST. Using the recorded information from this step, we will eliminate the one-hot encoding $e_i$ in the next step.

\begin{PRAalgorithm}[BST recording with $\mathsf{MaF}$.]{alg:RBST}
\renewcommand{\algorithmicrequire}{\textbf{Input:}}
\renewcommand{\algorithmicensure}{\textbf{Output:}}
    \Require $\sum^{d-1}_{i=0}\alpha_i\underbrace{\ket{q_i}}_{A}\underbrace{\ket{e_i}}_{B}\underbrace{\ket{0}^{\otimes 2d}}_{C}\underbrace{\ket{0}^{\otimes d}}_{D}$
\begin{spacing}{1.5}
    \Ensure $\sum^{d-1}_{i=0}\alpha_i\ket{q_i}\ket{e_i}\bigotimes_{k=1}^{d-1}\ket{f(q_i,k)}\ket{0}^{\otimes d}$
\end{spacing}

\State firstBranch $\gets$ true
\State k $\gets$ 1
\For{$j \gets 0$ to $n-1$}
    \If{firstBranch}
        \State firstBranch $\gets$ false
        \State X(C(0))
        \State CSWAP(A(j),C(0),C(1))
    \ElsIf{$\text{b(j)}\neq 0$}
        \For{$i \gets 1$ to $\text{b(j)}$}
            \If{$(\text{k}+i)$th branch node in its parent branch node's left subtree}
                \State CNOT(C(2PB(k+i)-2),C(2(k+i-1))
            \Else
                \State CNOT(C(2PB(k+i)-1),C(2(k+i-1))
            \EndIf
        \EndFor
        \State FANOUT(A(j),D(0:b(j)-1))
        \For{$i \gets 1$ to $\text{b(j)}$}
            \State CSWAP(D(i),C(2(k+i-1)),C(2(k+i)-1))
        \EndFor
        \State FANOUT(A(j),D(0:b(j)-1))
        \State k $\gets$ k$+$b(j)
    \EndIf
\EndFor
\end{PRAalgorithm}

\stepsection{Garbage elimination}\label{st4_2}
\indent
In this substep, we identify the value of $\ket{q_i}$ via the C register to flip the correct qubit in the B register. This can be done in parallel and requires only constant depth.

It is known that each one-hot encoding $\ket{e_i}$ corresponds to a state $\ket{q_i}$, and each $\ket{q_i}$ corresponds to a path from the root to a leaf in the BST. Suppose we want to eliminate the state $\ket{e_t}$.  The main approach is to apply a CNOT gate to flip the $t$th qubit in the B register. We can first identify the lowest branch node on the path corresponding to $\ket{e_t}$. If the path goes through the left subtree of this branch node, the control qubit of the CNOT gate is the first qubit of the two qubits recording this branch node; otherwise, it is the second qubit. When the B register is in the state $\ket{e_t}$, the control qubit will necessarily act, ultimately transforming $\ket{e_t}$ to $\ket{0}^{\otimes d}$. We can execute the process of eliminating $\ket{e_t}$ for all $t=0\sim d-1$ in parallel, because the control and target qubits for all the CNOT gates are distinct. It is obvious that the depth complexity is $O(1)$ and size complexity is $O(n)$ in this step. Algorithm~\ref{alg:Flip} presents the procedure for eliminating the garbage state. The notation LB(i) refers to the index of the lowest branch node on the path corresponding to $q_i$ in the BST.

\begin{PRAalgorithm}[Garbage Elimination.]{alg:Flip}
\renewcommand{\algorithmicrequire}{\textbf{Input:}}
\renewcommand{\algorithmicensure}{\textbf{Output:}}
    \Require $\sum^{d-1}_{i=0}\alpha_i\ket{q_i}\ket{e_i}\bigotimes_{k=1}^{d-1}\ket{f(q_i,k)}\ket{0}^{\otimes d}$
    \begin{spacing}{1.5}    
    \Ensure $\sum^{d-1}_{i=0}\underbrace{\alpha_i\ket{q_i}}_{A}\underbrace{\ket{0}^{\otimes d}}_{B}\underbrace{\bigotimes^{d-1}_{k=1}\ket{f(q_i,k)}}_{C}\underbrace{\ket{0}^{\otimes d}}_{D}$
    \end{spacing}

\For{$i \gets 0$ to $d-1$}
    \State Find the lowest branch node LB on the path corresponding to $\ket{e_i}$
    \If{the path go through the left subtree of LB}
        \State CNOT(C(2LB($i$)-2),B($i$))
    \Else
        \State CNOT(C(2LB($i$)-1),B($i$))
    \EndIf
\EndFor
\end{PRAalgorithm}
\stepsection{Reset C register}\label{st4_3}
\indent
As soon as the garbage state is eliminated, the C register can be rapidly deallocated. The method is similar to Step~\hyperref[st4_1]{4.1}. Additionally, the size and depth complexity is the same as in Step~\hyperref[st4_1]{4.1}.

After completing Step~\hyperref[Step4]{4}, the garbage state is eliminated and the desired state $\ket{\phi}$ is obtained. Without using $\mathsf{MaF}$, the process requires a circuit depth of $O(n\log d)$, while with $\mathsf{MaF}$, the depth can be reduced to $O(n)$. The size complexity is $O(d)$ in both cases.

\subsection{Proof}

Having introduced the four steps of the SQSP algorithm, we now integrate these procedures to derive and formally prove the overall complexity results. Specifically, we summarize and prove Theorem~\ref{th:1} and Theorem~\ref{th:2}, which correspond to the cases without and with the $\mathsf{MaF}$ circuit, respectively.
\begin{Proof of Theorem}[SQSP algorithm]
\label{pr1}
The SQSP algorithm begins with the input state $\ket{0}^{\otimes n}$. In the Step~\hyperref[Step1]{1}, a $\lceil\log d\rceil$-qubit state is prepared using a low-depth GQSP algorithm that encodes the target amplitudes. This step requires a circuit size of $O(d)$ and a depth of $O(\log d)$.

The Step~\hyperref[Step2]{2} constructs a one-hot state, enabling parallel execution of the subsequent permutation step. This is achieved by applying CSWAP gates to move the $\ket{1}$ to the correct position. To reduce the circuit depth, ancilla qubits and the copy operations are used to parallelize the CSWAP operations. As a result, the circuit depth is compressed to $O(\log^2 d)$ with total circuit size $O(d)$.

In the Step~\hyperref[Step3]{3}, OR-C-X gates are controlled in parallel by the one-hot states corresponding to all basis states, encoding each qubit sequentially to its correct value. This permutation step realizes the target state with a circuit size of $O(dn)$ and depth $O(n \log d)$. 

Finally, the Step~\hyperref[Step4]{4} eliminates the one-hot encoding. This is done by first constructing a BST of the target state and then recording the branch node information based on the BST. Similar to second step, a copy operation is used here to reduce the circuit depth. The CNOT gates are then applied to reset the one-hot state in parallel. The circuit size is $O(dn)$ and depth $O(n\log d)$ in this step.

By combining these four steps, an arbitrary $n$-qubit $d$-sparsity quantum state can be efficiently prepared, with a total circuit size of $O(dn)$, depth $O(n \log d)$ and $O(d)$ ancilla qubits.\hfill$\blacksquare$
\end{Proof of Theorem} 

\begin{Proof of Theorem}[SQSP algorithm with a $\mathsf{MaF}$ circuit]
\label{pr2}
The structure of this algorithm largely follows the same steps as in Prf.~\ref{pr1} and achieves the same circuit size and ancilla complexity. We highlight here the key differences that arise when using a $\mathsf{MaF}$ circuit, which allows fan-out operations to be implemented in constant depth and thus reduces the overall depth complexity.

The Step~\hyperref[Step1]{1}, GQSP, remains unchanged, with circuit size and depth being $O(d)$ and $O(\log d)$, respectively.

Since fan-out gates can perform the copy operation, the copy circuits in both Step~\hyperref[Step2]{2} and Step~\hyperref[Step4]{4} can be replaced with constant-depth fan-out gates. This reduces their depths from $O(\log^2 d)$ and $O(n \log d)$ to $O(\log d)$ and $O(n)$, respectively, while maintaining the same circuit sizes of $O(d)$ and $O(dn)$.

In Step~\hyperref[Step3]{3}, under the assumption that the input is a one-hot state, the OR-C-X gates behave equivalently to Par-C-X gates. Since Par-C-X gates can be implemented using Hadamard gates and fan-out gates, the circuit depth can be reduced from $O(n \log d)$ to $O(n)$, while maintaining the same circuit size of $O(dn)$.

In summary, the overall circuit depth can be reduced from $O(n \log d)$ to $O(n)$, making it independent of the sparsity $d$, while maintaining a circuit size of $O(dn)$ and $O(d)$ ancilla qubits.\hfill$\blacksquare$
\end{Proof of Theorem}
\vspace{-10pt}
\section{Conclusion}\label{chap:ccl}

In this work, we propose two efficient algorithms to prepare arbitrary sparse quantum state, with a focus on optimizing circuit depth under a limited number of ancilla qubits. The core idea is to first generate the correct amplitudes, and then relocate each amplitude to its designated position. In the first algorithm, we prepare each qubit sequentially rather than constructing the full basis states one by one. By using $O(d)$ ancilla qubits, we leverage copy operations and one-hot encoding to parallelize the computation. As a result, the overall circuit has size $O(dn)$, depth $O(n \log d)$, and requires $O(d)$ ancilla qubits. To further reduce the circuit depth, the second algorithm incorporates measurement and feedforward. By performing certain operations conditionally based on mid-circuit measurement outcomes, the circuit depth can be reduced to $O(n)$ while maintaining the same circuit size and ancilla requirements. 

For future work, it would be interesting to explore approaches that offer different trade-offs between ancilla qubits and circuit depth. Additionally, it is worth investigating more efficient state preparation algorithms adapted to different scaling of sparsity $d$ with respect to the number of qubits $n$. Another open question is to establish lower bounds on the circuit depth required for quantum state preparation under measurement-and-feedforward (MaF) architectures.

\section*{Acknowledgement}
This work is supported by NSTC QC project under Grant no. 114-2119-M-001-002- and 113-2112-M-007-043-.

\section*{APPENDIX: LOCAL OPERATION AND CLASSICAL COMMUNICATION}

Many studies on LOCC and its related applications have been published over the past decade.~\cite{LOCC_2014,LOCC_2024,LOCC_2025} In the following, we present an introduction to LOCC~\cite{LOCC} and LAQCC~\cite{shallow_2024}, and provides a detailed comparison with $\mathsf{MaF}$.

Local operation and classical communication (LOCC) is a fundamental concept in quantum information theory used to describe operations on distributed quantum systems under specific constraints. It consists of two key components:
\begin{enumerate}[label=(\arabic*)]
\item \textbf{Local operations} Each party can only perform operations on their local quantum subsystem.  These operations include measurements, unitary transformations, additions of ancilla, and discarding or preparing new quantum states.
\item \textbf{Classical communication} Parties can exchange measurement outcomes through classical channels but cannot directly share quantum information, such as quantum states or entanglement. The purpose of classical communication is to coordinate their actions such that another local operation is performed, conditioned on the information received.
\end{enumerate}
Inspired by the concept of LOCC in quantum information, an LOCC circuit architecture is proposed, where the circuit is constructed with geometrically local gates, and these local gates are controlled by classical information computed from previous measurement results. Such a circuit is referred to as an LOCC-circuit.

Local alternating quantum classical computations (LAQCC), similar to LOCC, is a architecture that incorporates mid-circuit measurements and real-time feedforward operations. In this framework, quantum and classical operations are alternated in a controlled manner,and each qubit arranged in a grid can only interact with their immediate neighbors. The main distinction between LOCC and LAQCC is that LAQCC restricts quantum operators to constant depth, classical operators to logarithmic depth, and allows at most a constant number of alternations between them. From this, we can conclude that any LAQCC-circuit has constant total quantum depth.

Since the LOCC-circuits also incorporate the mid-circuit measurements, they fall within the category of circuits with $\mathsf{MaF}$. The key difference is that, under the LOCC framework, each quantum gate is allowed to interact only with adjacent qubits, whereas circuits with $\mathsf{MaF}$ are not subject to this constraint. We note that the fan-out gate shown in Fig.~\ref{fig:LAQCC} also operates within the LOCC framework. However, in this work, we utilize the fan-out gates where the control and target qubits are not necessarily adjacent. Moreover, other operations in the circuit are not strictly local either. As a result, the SQSP circuits we propose fall under the circuit with $\mathsf{MaF}$ category rather than being constrained to LOCC.

\bibliographystyle{apsrev4-2}
\bibliography{Sample}
\end{document}